\documentclass[aps,pra,twocolumn,amsmath,amssymb,superscriptaddress]{revtex4-1}

\usepackage{physics}
\usepackage{float}
\usepackage{graphicx}
\usepackage{dcolumn}
\usepackage{bm}
\usepackage{multirow}
\usepackage{color}
\usepackage{booktabs}
\usepackage{hyperref}

\graphicspath{{Figures/}} 

\begin{document}

\title{Sharp estimates for the integrated density of states in Anderson tight-binding models}

\author{Perceval Desforges}
\affiliation{
Laboratoire de Physique de la Matière Condensée, Ecole Polytechnique, CNRS, Institut Polytechnique de Paris, 91120 Palaiseau, France
}

\author{Svitlana Mayboroda}
\affiliation{
School of Mathematics, University of Minnesota, Minneapolis, Minnesota 55455, USA
}

\author{Shiwen Zhang}
\affiliation{
School of Mathematics, University of Minnesota, Minneapolis, Minnesota 55455, USA
}

\author{Guy David}
\affiliation{
Université Paris-Saclay, CNRS, Laboratoire de mathématiques d'Orsay, 91405, Orsay, France
}

\author{Douglas N. Arnold}
\affiliation{
School of Mathematics, University of Minnesota, Minneapolis, Minnesota 55455, USA
}

\author{Wei Wang}
\affiliation{
School of Mathematics, University of Minnesota, Minneapolis, Minnesota 55455, USA
}

\author{Marcel Filoche}
\affiliation{
Laboratoire de Physique de la Matière Condensée, Ecole Polytechnique, CNRS, Institut Polytechnique de Paris, 91120 Palaiseau, France
}


\date{\today}

\begin{abstract}
Recent work [G.~David, M.~Filoche, and S.~Mayboroda, Adv.~Math.\ (to be published)] has proved the existence of bounds from above and below for the integrated density of states (IDOS) of the Schr\"odinger operator throughout the spectrum, called the \emph{landscape law}. These bounds involve dimensional constants whose optimal values are yet to be determined. Here, we investigate the accuracy of the landscape law in 1D and 2D tight-binding Anderson models, with binary or uniform random distributions. We show, in particular, that in 1D, the IDOS can be approximated with high accuracy through a single formula involving a remarkably simple multiplicative energy shift. In 2D, the same idea applies but the prefactor has to be changed between the bottom and top parts of the spectrum.
\end{abstract}

\maketitle

%\tableofcontents

\section{Introduction}

In single-particle quantum systems subject to random potential, the integrated density of states [IDOS, or counting function, defined as the number of eigenvalues per unit volume smaller than a given energy and denoted hereafter as $N(E)$] departs significantly at low energy from the high energy asymptotic behavior known as Weyl's formula. According to the latter, in the absence of any potential, $N(E)$ scales as $E^{d/2}$ where $d$ is the ambient dimension. However, in the presence of a disordered or random potential the IDOS exhibits a very slowly growing tail at low energy. In 1964, Lifshitz proposed a model based on scattered impurities where the IDOS would drop off exponentially as $E$ approaches its minimum value $E_0$, forming what is known as a \emph{Lifshitz tail}~\cite{Lifshitz1964, Lifshitz1965}:
\begin{equation}\label{eq:Lifshitz_1}
N(E) \sim C \exp \left( -c (E-E_0)^{-\frac{d}{2}} \right)\,.
\end{equation}
Since then, understanding the precise behavior of the density of states in the presence of disorder has been the subject of a very rich literature (for an extended review on the topic, the reader can refer to Refs.~\cite{Lifshits1988, Pastur1992, Kirsch2007, Grabsch2014, Konig2016}). The existence of Lifshitz tails for the Poisson random potential was proved in Refs.~\cite{Benderskii1970, Donsker1975, Nakao1977, Pastur1977}. Later Kirsch and Martinelli gave a proof close to Lifshitz’s intuition for a large class of random potentials in the continuous setting in Ref.~\cite{Kirsch1983}, while Simon generalized the argument to the tight binding model~\cite{Simon1985}. These are only a few isolated results and we do not aim to provide an exhaustive list of the literature. It is important to mention however that there exist exact asymptotic results on Lifshitz tails for specific models~\cite{Simon1985, Mezincescu1986, Biskup2001, Konig2016}. Nonetheless, we are still lacking a general understanding for all models, and the only mathematical statement that could be rigorously proven in \emph{full generality} does not have the form of Eq.~\eqref{eq:Lifshitz_1} but rather the weaker form:
\begin{equation}\label{eq:Lifshitz_2}
\lim_{E \to E_0}  \frac{\ln(\abs{\ln(N(E))})}{\ln(E-E_0)} = -\frac{d}{2} \quad.
\end{equation}
These results are asymptotic in the limit of vanishing $E-E_0$. Away from the asymptotic behavior at low energy, Klopp and Elgart showed that in the weak disorder limit these Lifshitz tails extend roughly up to the average of the potential~\cite{Klopp2002a, Klopp2002b, Elgart2009}. To this day, many unsolved questions remain concerning these Lifshitz tails: (i)~Can one improve the known results by deriving a general estimate on $\ln(N(E))$ and not on $\ln(\abs{\ln(N(E))})$? (ii)~If so, how does disorder enter the estimate? For instance, can one quantify existing results showing logarithmic corrections for random uniform disorder as compared to binary disorder~\cite{Nieuwenhuizen1989, Politi1988}? (iii)~Can one derive a precise estimate for the full spectrum instead of asymptotic near the lower bound of the spectrum?

In this paper, we present a new function, denoted by $N_u(E)$ and called \emph{landscape law}~\cite{David2019landscape}, that provides estimates for the actual counting function (from above and below) throughout the spectrum. This function is obtained from the \emph{localization landscape}, a theoretical tool introduced in 2012 and developed in the recent years~\cite{Filoche2012, Arnold2016}. Not only do these estimates cover the entire range of energy for any type of potential or disorder in continuous models, but they also provide the asymptotic behavior of $\ln(N(E))$ at low energy for random uniform or binary disorder, thus removing a $\log$ from the previously known results. In particular, they recover the logarithmic correction in the case of the uniform Anderson model. We investigate numerically the optimal constants involved in the bounds, and observe their similarity for both binary and uniform Anderson models. Finally, we test whether these mathematically proven bounds from above and below could in fact be merged into one single approximate formula based on $N_u$, thus providing a very fast and efficient way of predicting the behavior of the IDOS on the entire spectrum even in a random or complicated system.

\section{The landscape law}

We consider a $d$-dimensional tight-binding model. The corresponding Hamiltonian is
\begin{equation}
\hat{H} = \sum_i V_i a_i^{\dagger} a_i - t \sum_{\langle i,j \rangle} \left( a_i^{\dagger} a_j + \text{H.c.} \right),
\end{equation}
where $\langle i,j \rangle$ denotes the sum over nearest neighbors, $t$ is the hopping term, $V_i$ is the on-site random potential (on a grid of lattice parameter 1), and $a_i^{\dagger}$ and $a_i$ are the creation and annihilation operators, respectively. From now on, $t$ will be taken equal to 1, thus setting the energy scale. The $V_i$ are i.i.d.~variables and follow a random law which can be either uniform or binary in [0, $V_{\max}$]. The localization landscape $u$ in this system is defined as the solution to $\hat{H} u=1$, the right-hand side being the constant vector. To ensure that the landscape is positive everywhere, the potential $V$ is uplifted by a quantity $2d$, where $d$ is the embedding dimension. Consequently, the lowest bound $E_0$ of the spectrum in all subsequent examples is $E_0=0$ and the spectrum lies in the interval [0, $V_{\max} + 4d$]. It has been shown in Ref.~\cite{Lyra2015} that the function $W \equiv 1/u$ defines an effective potential for all quantum states in the tight-binding model, and that this potential provides a remarkably accurate estimate of the energy of the lower-energy states.

Using this effective potential, the function $N_u(E)$ is defined as follows: for a given energy $E$, we partition the entire domain into $d$-cubes (intervals in 1D, squares in 2D,\ldots) of side length $E^{-1/2}$. $N_u(E)$ is then defined as the fraction of such cubes for which the minimum of $W$ over the cube is smaller than $E$:
\begin{align}\label{eq:landscape_law}
N_u(E) \equiv \frac{1}{\abs{\Omega}} \times \Big( & \textrm{ number of cubes of size } \frac{1}{\sqrt{E}} \nonumber \\
& \textrm{ where } \min(W) \leq E \Big) \,.
\end{align}
For the continuous model, it has been mathematically proven in Ref.~\cite{David2019landscape} that there exist constants $C_4$, $C_5$, $C_6$ such that $N_u$ satisfies the following inequalities:
\begin{equation}\label{eq:inequalities}
 C_5 N_u \left( C_6 E \right) \leq N(E) \leq N_u \left( C_4 E \right),
\end{equation}
where $N(E)$ is the IDOS per unit volume. The constants $C_5$ and $C_6$ depend only on the dimension~$d$ and on the average of the potential, and $C_4$ depends only on the dimension. When the potential is random, this inequality is verified for the expectations of the IDOS (note that these expectations become finite deterministic quantities in the limit of an infinite domain). These inequalities are universal bounds for the counting function $N(E)$ of a Schr\"odinger Hamiltonian throughout the entire spectrum. In other words, unlike Weyl's formula or Lifshitz tails, they are not asymptotic. The proof is rather technical and is based on the analysis of the low values of the effective potential $W$. A sketch of the proof is given in the Supplemental Material. We are currently preparing a version of this proof for discrete tight-binding models~\cite{Arnold2021}.

\begin{figure}[ht]
\includegraphics[width=0.45\textwidth]{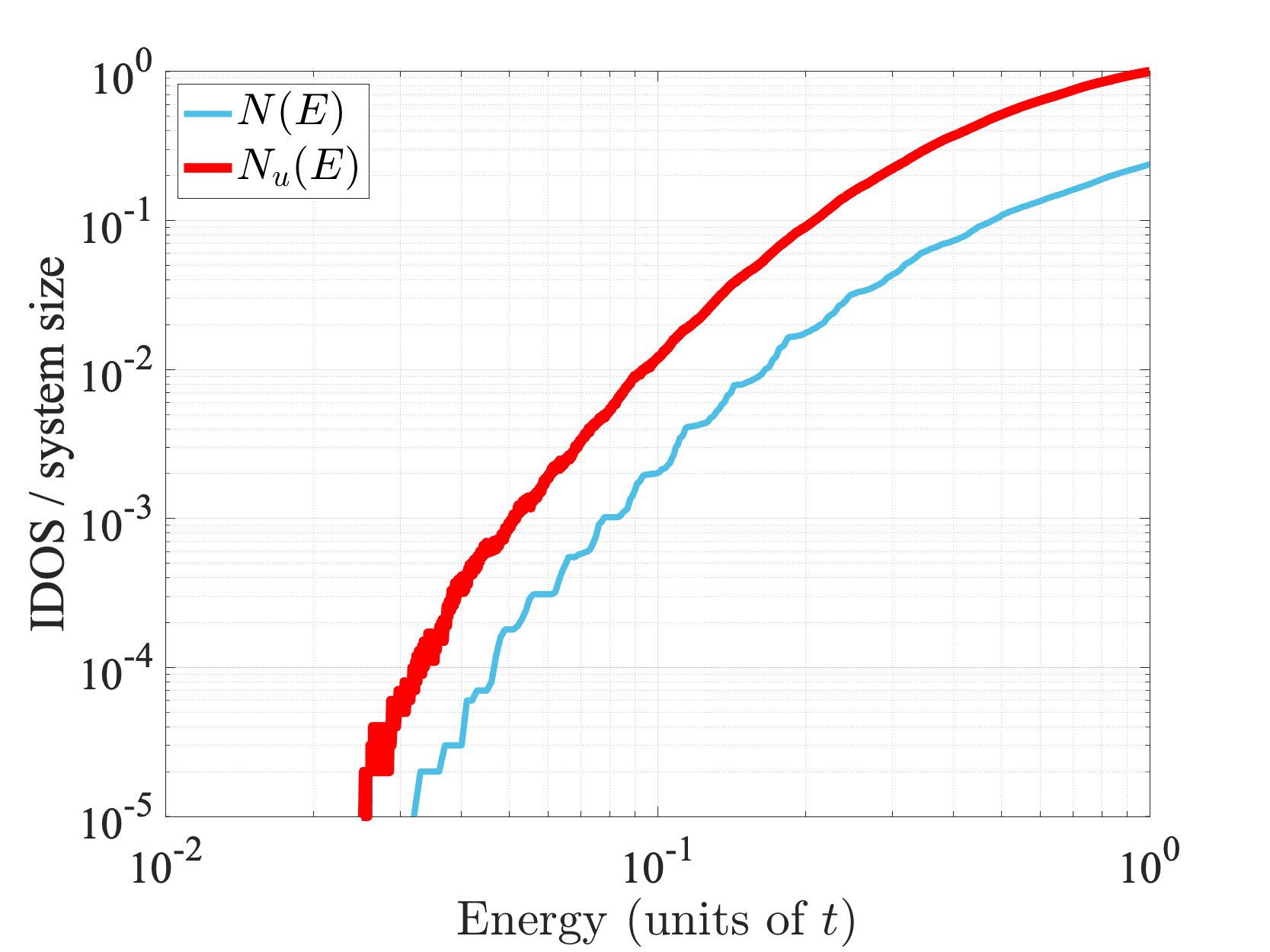}
\caption{Counting function $N(E)$ (blue) and landscape law $N_u(E)$ (thick upper red) for one realization of a one-dimensional binary Anderson tight-binding model. The number of sites is $N=10^5$ and the values of the on-site potential are either 0 or 1.}
\label{fig:1D_binary}
\end{figure}

An example of the sharpness and the predictive power of this inequality is provided in Fig.~\ref{fig:1D_binary} which displays the actual IDOS $N(E)$ (blue) and the landscape law $N_u(E)$ (red) for one realization of a random i.i.d.\,binary disorder with periodic boundary conditions. The potential can take the values either 0 or $V_{\max} = 1$ with equal probability on each site of a one-dimensional (1D) domain of $N=10^5$~sites. $N(E)$ is computed using the $LDL^T$ factorization and Sylvester's law of inertia~\cite{Parlett1998}. One can see how the two curves, plotted on a log-log scale, follow each other very closely. On this log-log plot, the upper and lower bounds of \eqref{eq:inequalities} would correspond simply to horizontal and vertical translations of the graph of $N_u(E)$.

While~Ref.~\cite{David2019landscape} proves the existence of constants $C_4$, $C_5$, $C_6$ fulfilling~\eqref{eq:inequalities}, it does not bring any insight into their sharpest values. Indeed, strictly speaking, Ref.~\cite{David2019landscape} gives a tube containing the IDOS (in log-log plot), and while it is remarkable that the tube diameter does not depend on the energy, it could be quite wide if the constants are very different. The goal of this paper is threefold: first, to demonstrate the accuracy of the landscape law in approximating the actual IDOS. Second, we indicate how to determine numerically the sharpest values for the constants entering the bounds in~\eqref{eq:inequalities}. This is of particular relevance for $C_4$ which is predicted to be universal, i.e., to depend only on the dimension~$d$ and not on the particular potential. Third, we assess the possibility of providing an optimal approximation to the IDOS $N(E)$ (rather than a tube), i.e., to find a constant $C_{5,\textrm{fit}}$ such that $N(E) \approx C_{5,\textrm{fit}} N_u(C_6 E)$. 

\begin{figure}[H]
\center
\includegraphics[width=0.46\textwidth]{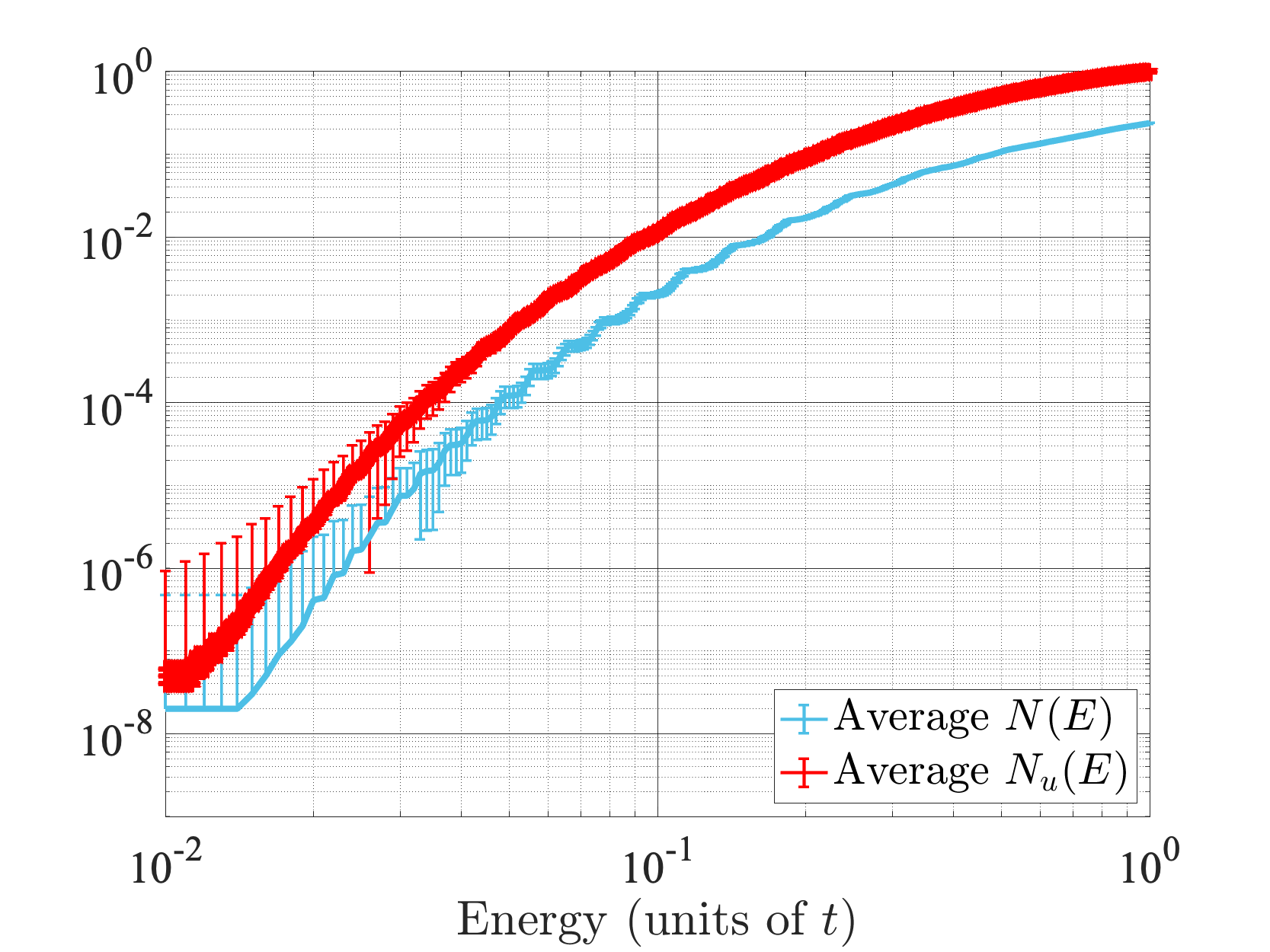}
\\(a)\\
\includegraphics[width=0.23\textwidth]{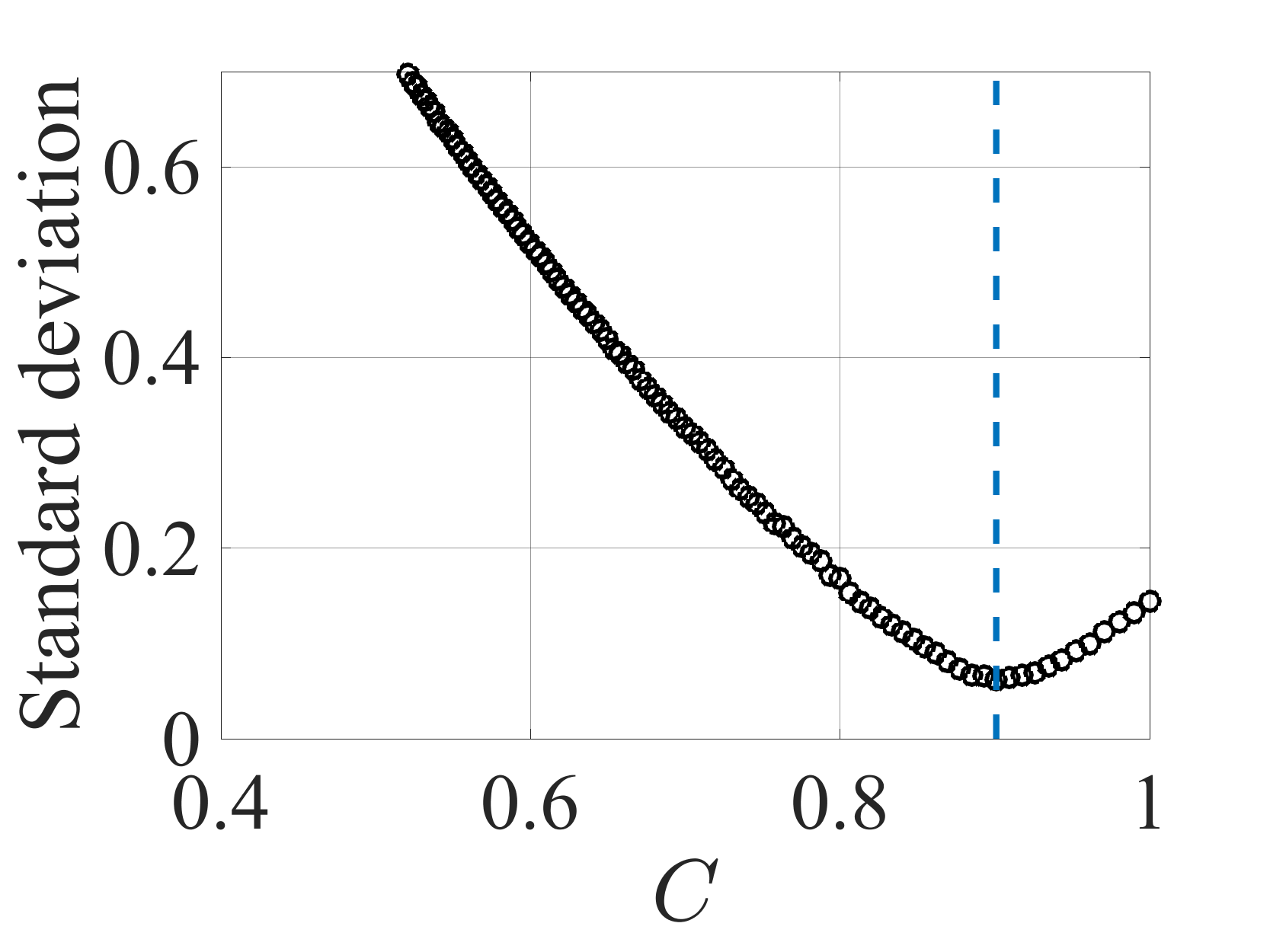}
\includegraphics[width=0.23\textwidth]{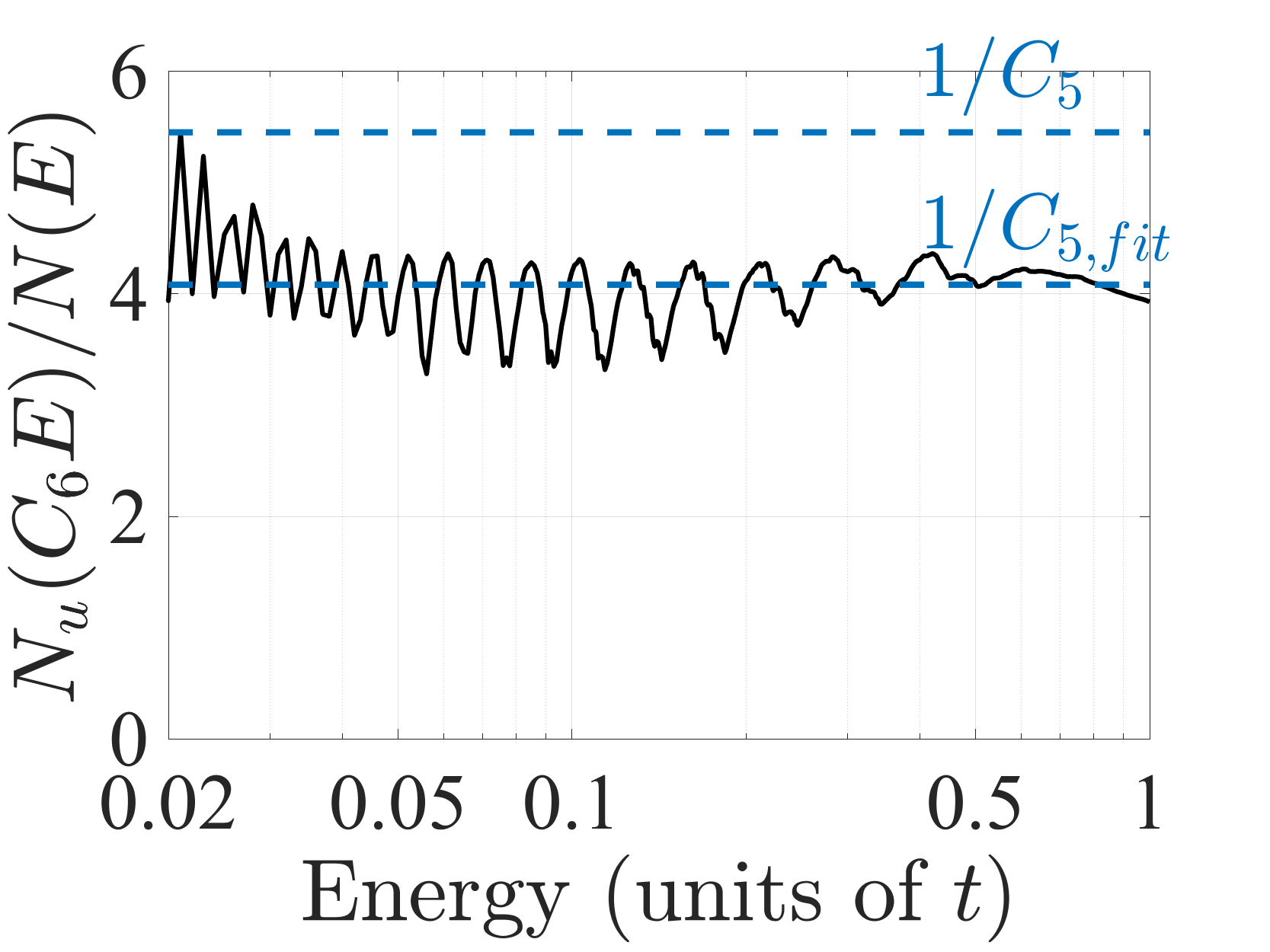}
\\ \hskip 3mm (b) \hskip 36mm (c)\\
\includegraphics[width=0.46\textwidth]{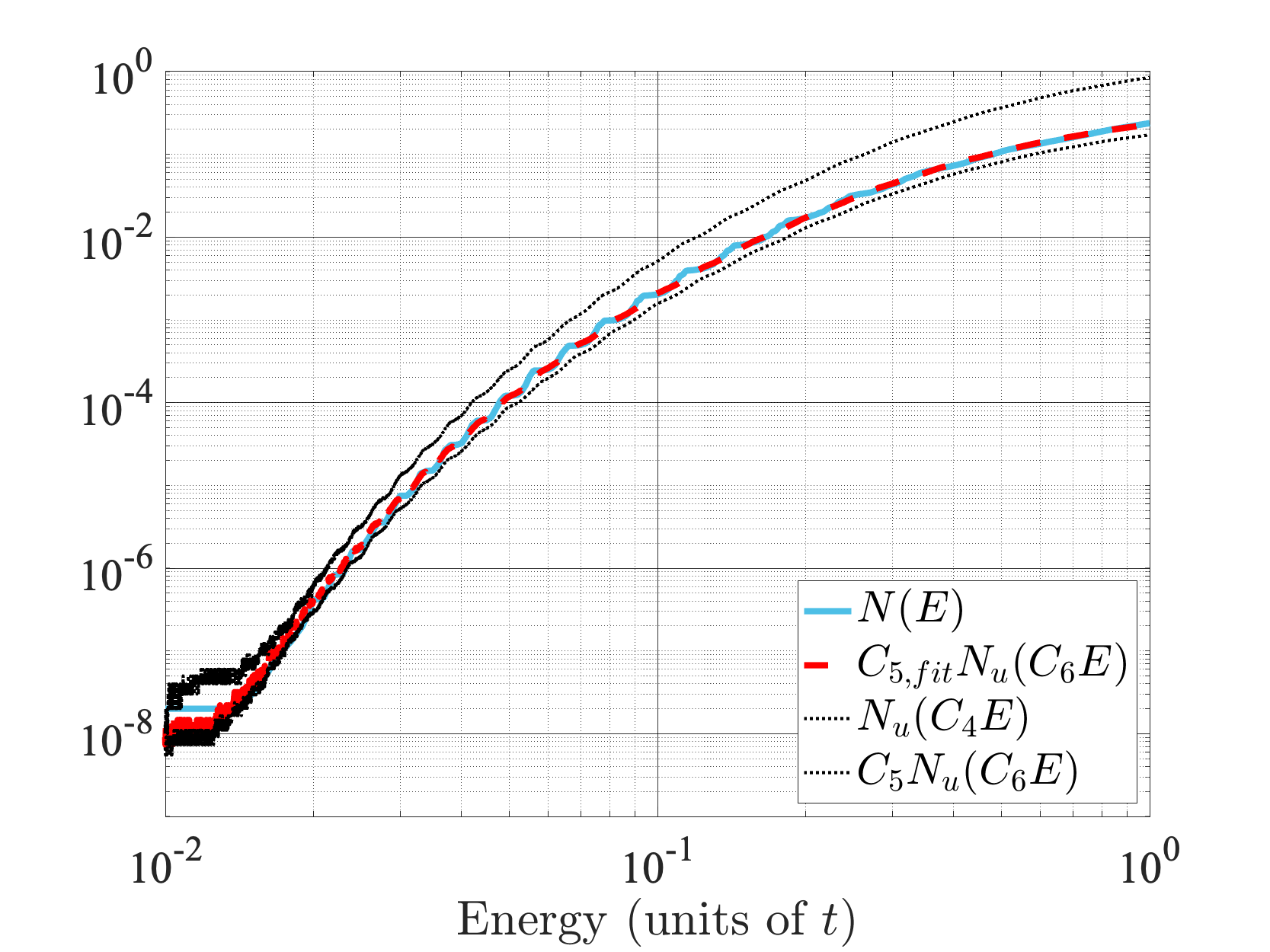}
\\(d)\\
\caption{(a) $N(E)$ (blue) averaged over 1000~random realizations, and averaged landscape law $N_u(E)$ (thick upper red) for a one-dimensional binary Anderson tight-binding model of size $N=10^5$ and $V_{\max} = 1$. The error bars show the standard deviation over the 1000 realizations (the bottom bars are not displayed when they are larger than the value itself, i.e., when they cross the horizontal axis). (b) Standard deviation of the distribution of values of $\ln(N_u(CE)/N(E))$ for various values of $C$. The minimum around $C=0.90\approx1/1.11$ provides the value of $C_6$. (c) Plot of $N_u(C_6 E)/N(E)$. The maximum shows that one can take $C_5=1/5.45$. A best fit for $N(E)$ is obtained by taking the average value $C_{5,\textrm{fit}} \approx 1/4.08$. (d) Final comparison between the original $N(E)$ (blue), the best fit $C_{5,\textrm{fit }} N_u(C_6 E)$ (dashed red), and the two bounds from above and below in Eq.~\eqref{eq:inequalities} $N_u(C_4E)$ and $C_5N_u(C_6E)$ (dotted lines). Note that the best fit is so close to the actual IDOS that the blue line is almost invisible.}
\label{fig:1D_binary_average}
\end{figure}

\subsection{1D systems: results}

One starts with the same system as in Fig.~\ref{fig:1D_binary}, but this time $N(E)$ is averaged over 1000~realizations. Figure~\ref{fig:1D_binary_average}(a) displays the corresponding $N(E)$ and $N_u(E)$ together with their standard deviations represented as error bars (the bottom bars are not displayed when they are larger than the value itself, i.e., when they cross the horizontal axis). To determine the constants $C_4$, $C_5$, $C_6$, we first restrict our study to the domain $E>0.02$ to avoid the noise at very low energy. We observe that the graph of $N_u(E)$ is always above the graph of $N(E)$, which means that $C_4 < 1$. This fact derives from the definition of $N_u$, and we will discuss it further down. The value of the constant $C_4$ corresponds in log-log scale to the largest possible right-shift of the graph of $N_u$ (or, in other words, to the smallest possible value of $C$) such that $N(E) \le N_u(CE)$. Here, this value is found to be $C_4 \approx 0.79$ (or $1/C_4 \approx 1.26$). To find the values of $C_5$ and $C_6$, we first look for the optimal value $C$ such that $N(E)/N_u(CE)$ is as constant as possible. This is achieved by taking the minimum of the standard deviation of $\ln(N(E)/N_u(CE))$ when varying $C$. Figure~\ref{fig:1D_binary_average}(b) displays this standard deviation for values of $C$ ranging from 0.5 to 1. One observes a clear minimum at $C_6 \approx 0.90$ (or $1/C_6 \approx 1.11$). Finally, the minimum of the graph of $N(E)/N_u(C_6 E)$ provides us the sharpest value of $C_5$ [see Fig.~\ref{fig:1D_binary_average}(c)]: It is here $C_5 \approx 0.18$. However, one can observe that if we were looking for a best fit for $C_5$ [instead of a lower bound for~\eqref{eq:inequalities}], then the best fit would be closer to $C_{5,\textrm{fit }} \approx 1/4.08$ (obtained by computing the average of $\ln(N(E)/N_u(C_6 E))$ for $E>0.02$). With these constants, the agreement between the actual IDOS and the rescaled formula based on the localization landscape is excellent throughout the computed spectrum [see Fig.~\ref{fig:1D_binary_average}(d)]. This means that the inequalities in \eqref{eq:inequalities} can almost be transformed into an equality:
\begin{equation}\label{eq:landscape_fit}
N(E) \approx C_{5,\textrm{fit} } \, N_u(C_6 E) \,.
\end{equation}

\begin{figure}[h!]
\includegraphics[width=0.46\textwidth]{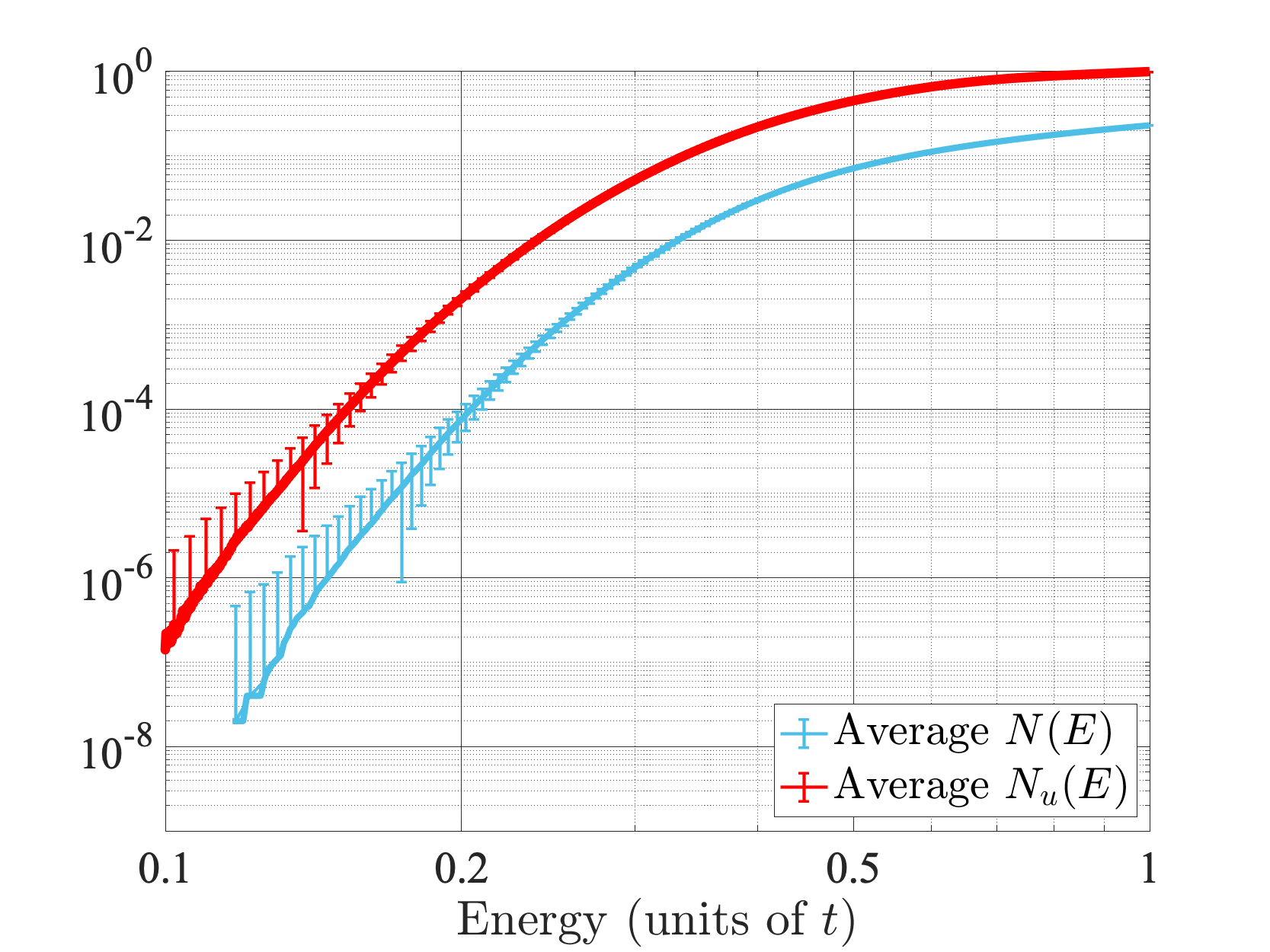}
\\(a)\\
\includegraphics[width=0.23\textwidth]{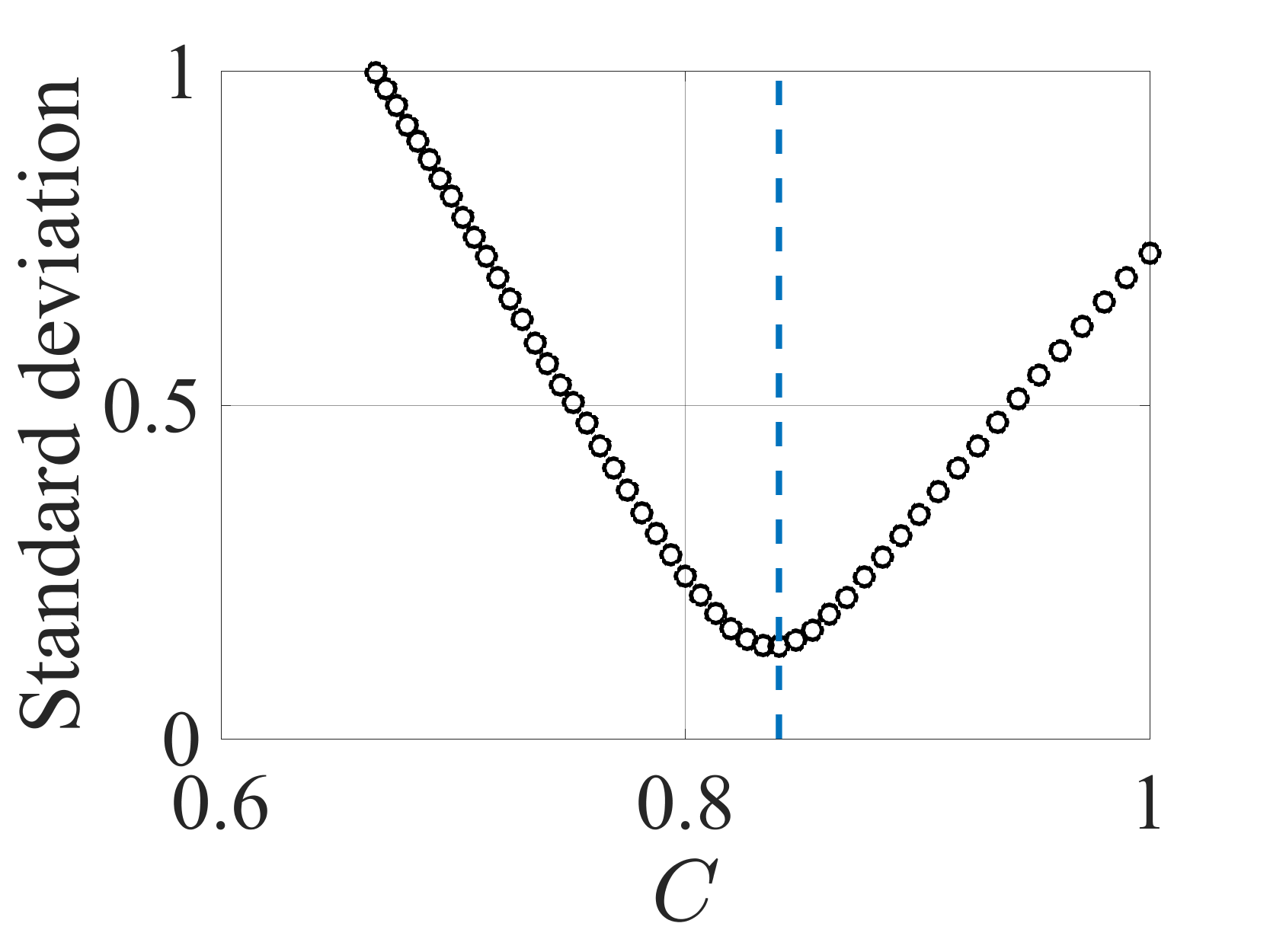}
\includegraphics[width=0.23\textwidth]{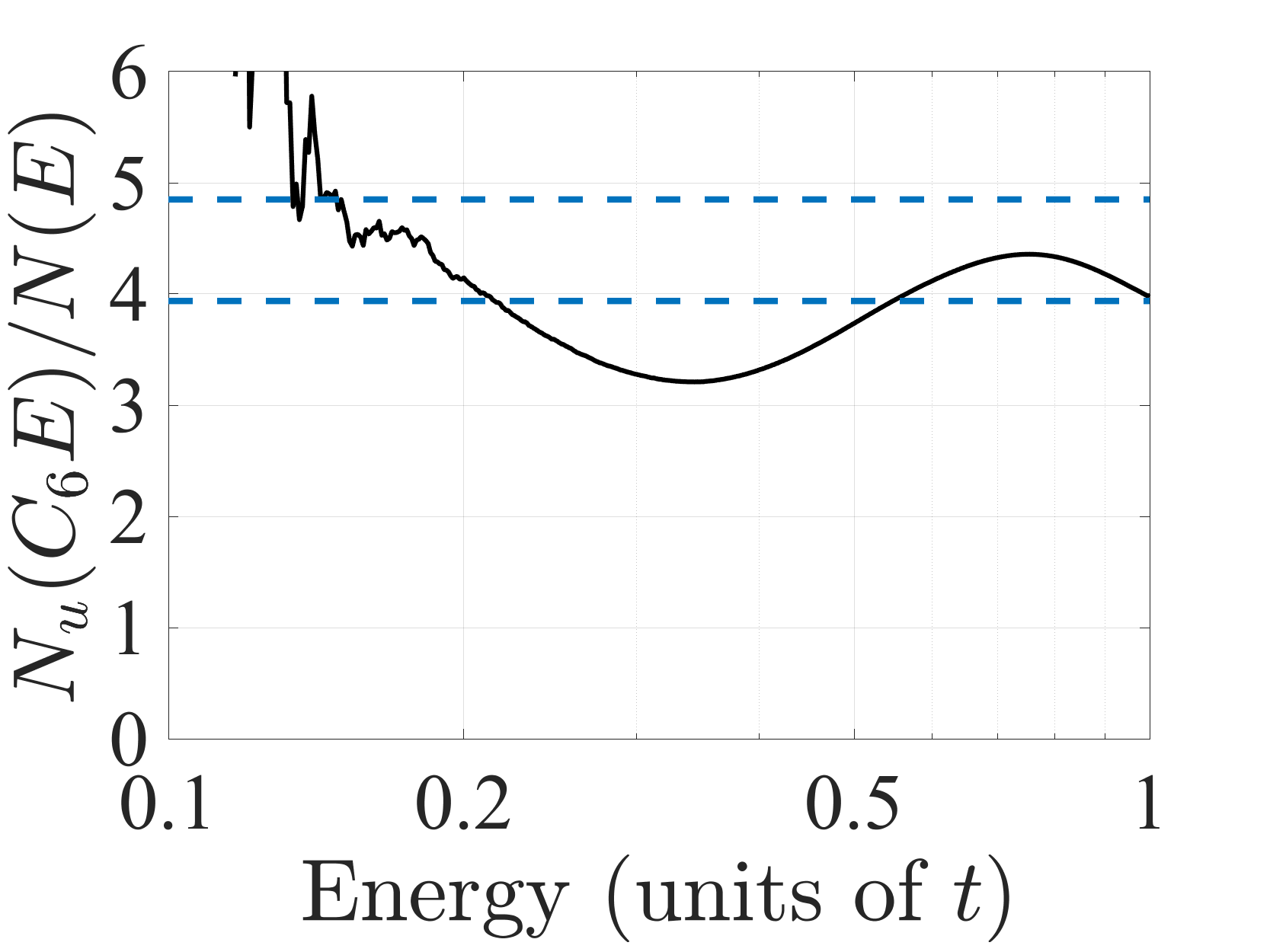}
\\ \hskip 3mm (b) \hskip 36mm (c)\\
\includegraphics[width=0.46\textwidth]{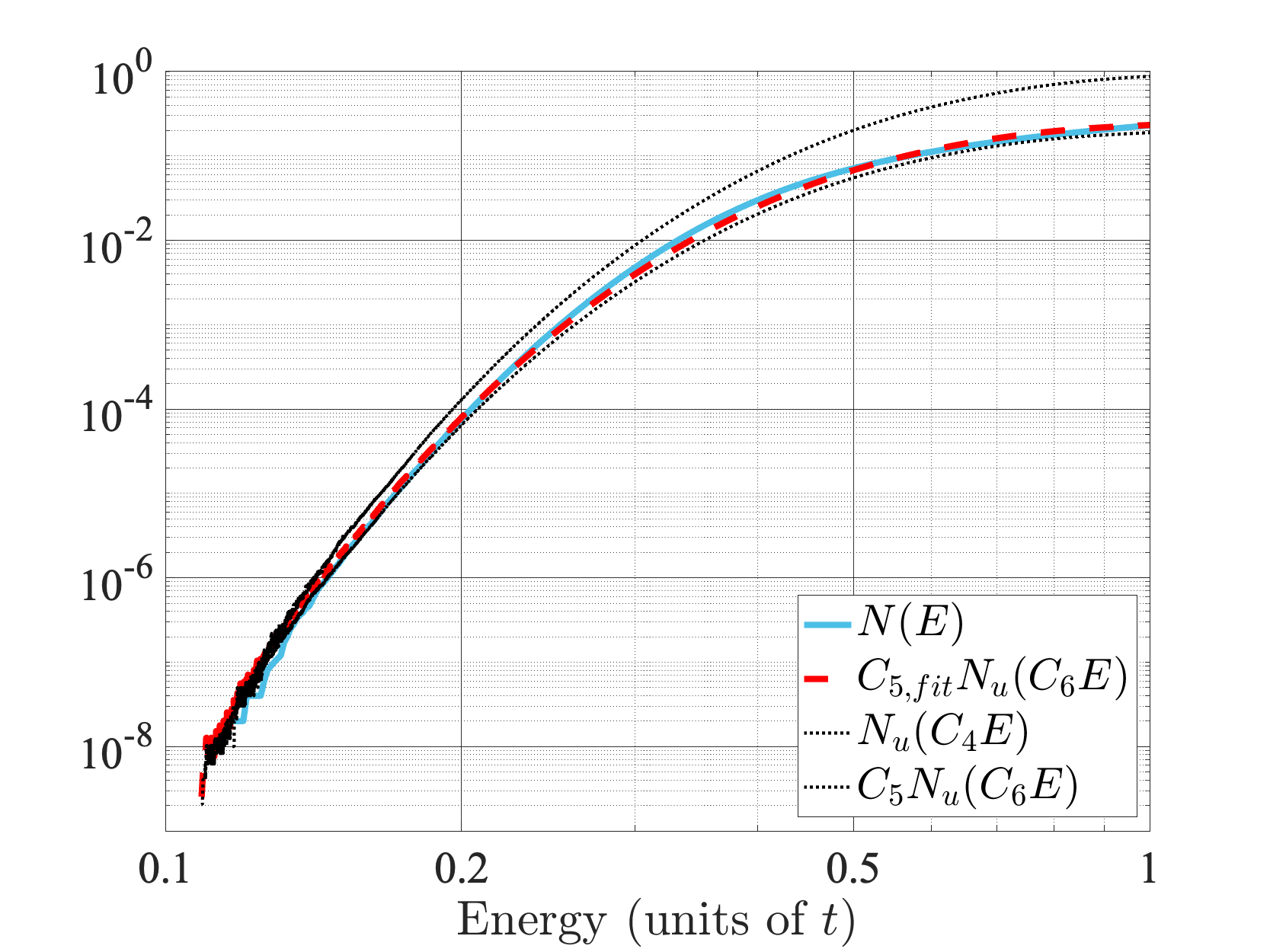}
\\(d)\\
\caption{(a) $N(E)$ (blue) averaged over 1000~random realizations, and averaged landscape law $N_u(E)$ (thick upper red), for a one-dimensional uniform Anderson tight-binding model of size $N=10^5$ and $V_{\max} = 1$. (b) Standard deviation of the distribution of values of $\ln(N_u(CE)/N(E))$ for various values of $C$. The minimum around $C=0.84\approx 1/1.19$ provides the value of $C_6$. (c) Plot of $N_u(C_6 E)/N(E)$. The maximum shows that one can take $C_5=1/4.85$. A best fit for $N(E)$ is obtained by taking the average value $C_{5,\textrm{fit }} \approx 1/3.94$. (d) Final comparison between the original $N(E)$ (blue), the best fit $C_{5,\textrm{fit }} N_u(C_6 E)$ (dashed red), and the two bounds from above and below $N_u(C_4E)$ and $C_5N_u(C_6E)$ (dotted lines).}
\label{fig:1D_uniform_average}
\end{figure}

The same methodology is then applied to a 1D uniform Anderson tight-binding model ($N=10^5$), and to two-dimensional (2D) binary and uniform Anderson tight-binding models. Figure~\ref{fig:1D_uniform_average} displays the results for a uniform Anderson model of disorder amplitude $V_{\max}=1$. Here also, we observe that the landscape law $N_u(E)$ follows very closely the actual IDOS $N(E)$. After computation, the value found for $C_4$ in this case is $C_4 \approx 0.78= 1/1.28$. Further analysis of the standard deviation of the values of $N_u(C E)/N(E)$ as a function of $C$ [see Fig.~\ref{fig:1D_uniform_average}(b)] yields $C_6 \approx 0.84=1/1.19$. Plotting now $N_u(C_6 E)/N(E)$ as a function of $E$ ([Fig.~\ref{fig:1D_uniform_average}(c)], one observes that it oscillates slowly between 3 and 5 in the noiseless part of the graph. A possible choice for $C_5$ is then $C_5 = 1/4.85$, but the best fit is obtained for $C_{5,\textrm{fit }}=1/3.94$, as confirmed in Fig.~\ref{fig:1D_uniform_average}(d).

\begin{figure}[h!]
\includegraphics[width=0.46\textwidth]{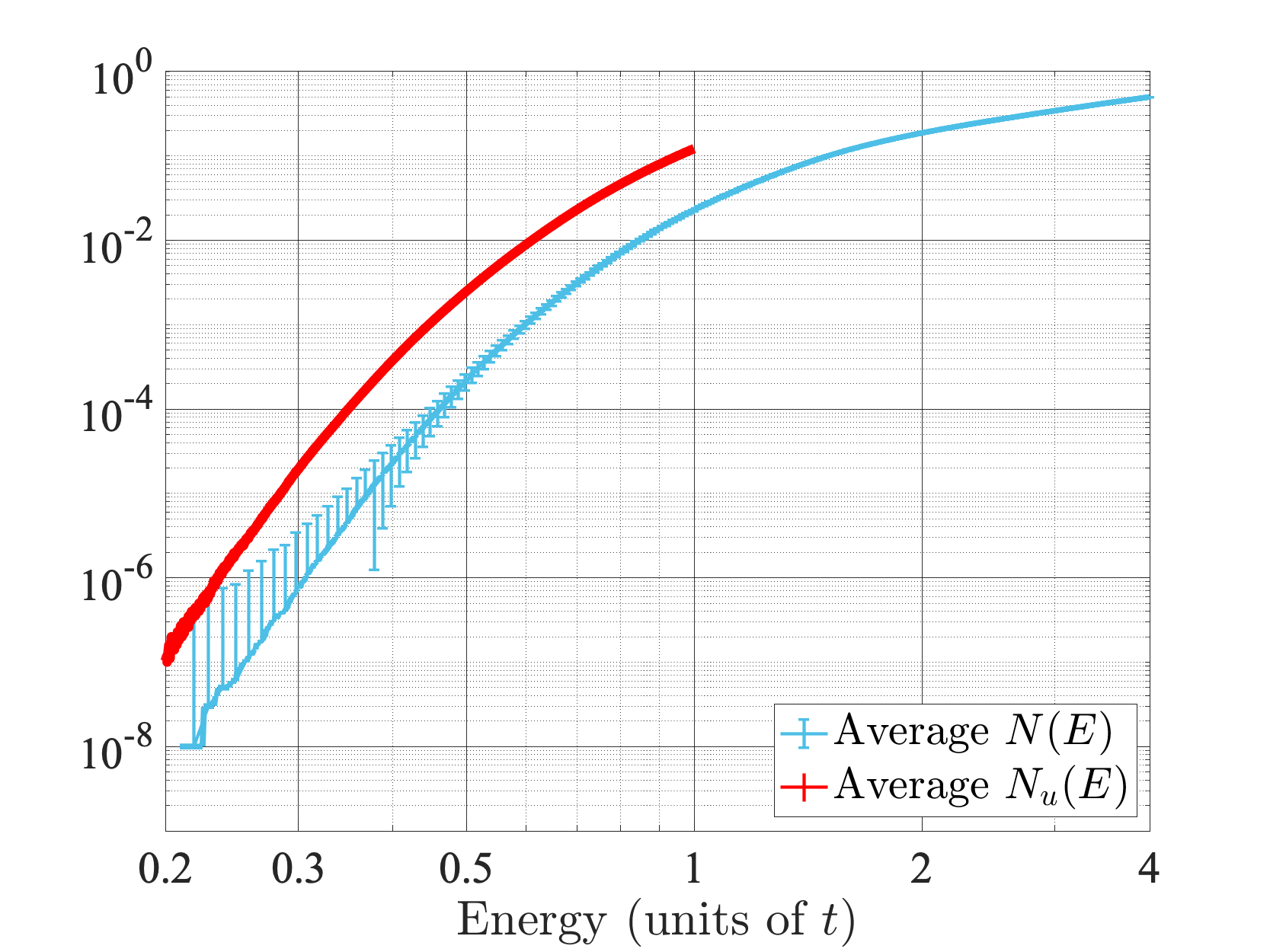}
\\(a)\\
\includegraphics[width=0.23\textwidth]{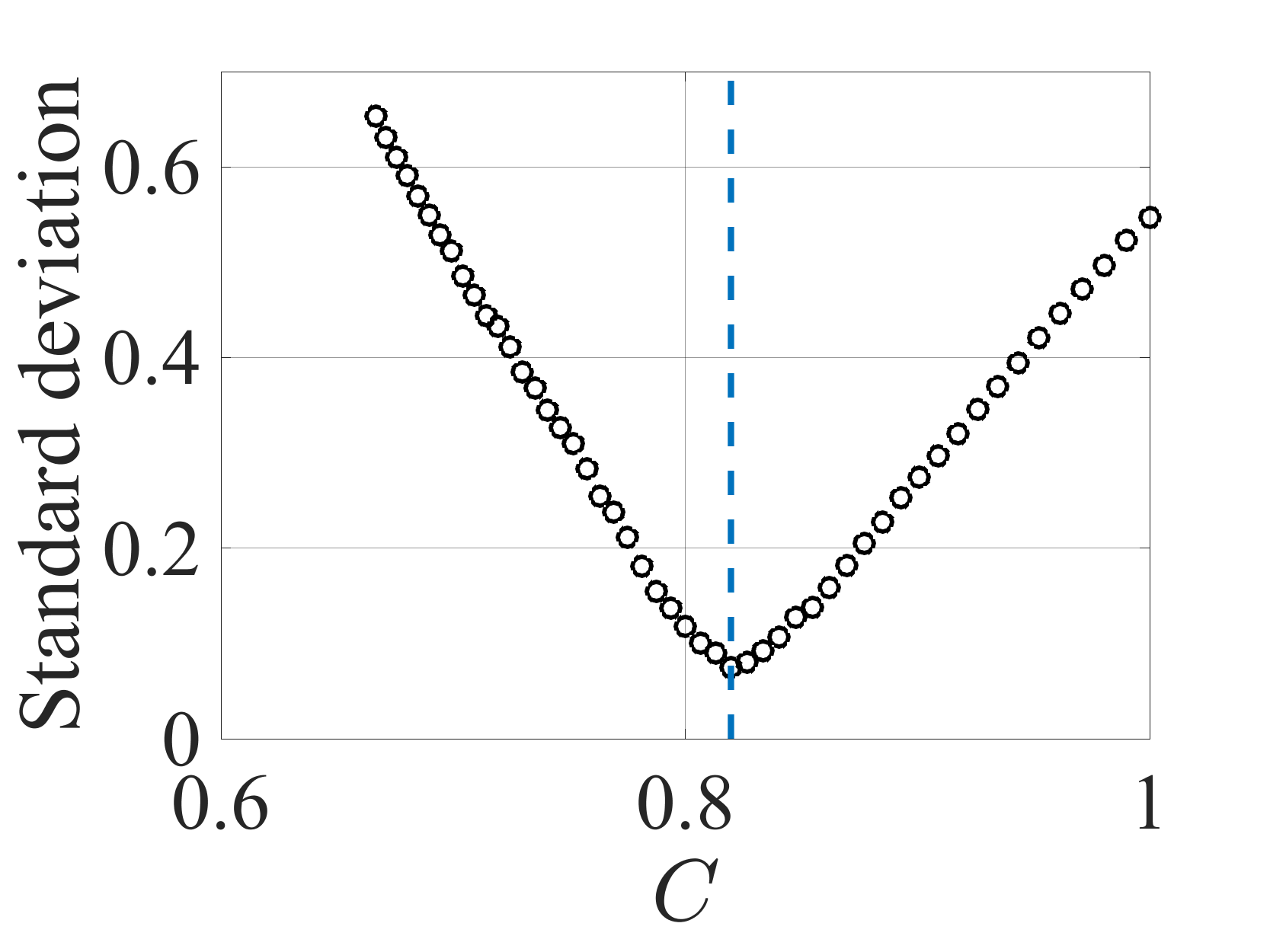}
\includegraphics[width=0.23\textwidth]{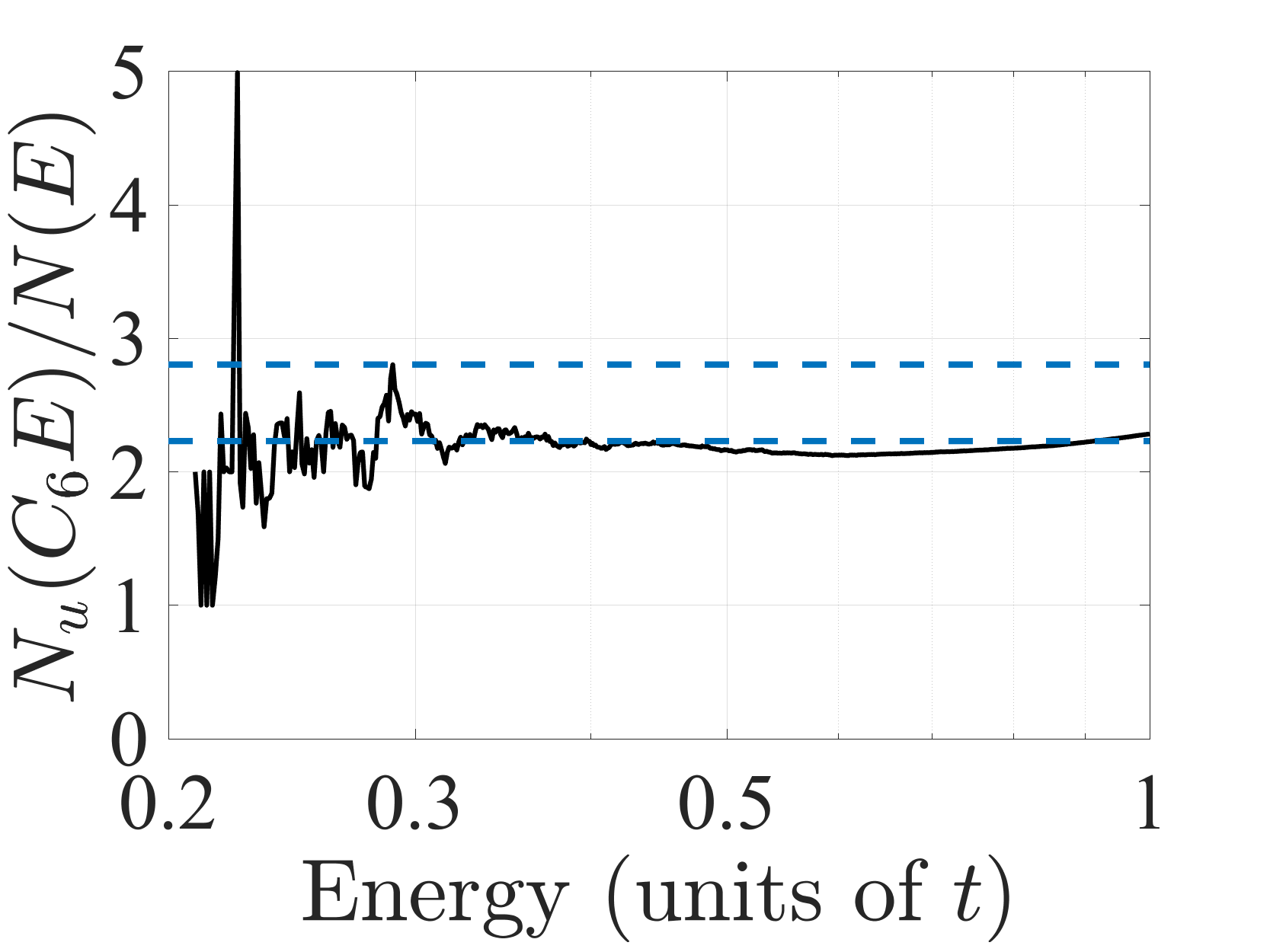}
\\ \hskip 3mm (b) \hskip 36mm (c)\\
\includegraphics[width=0.46\textwidth]{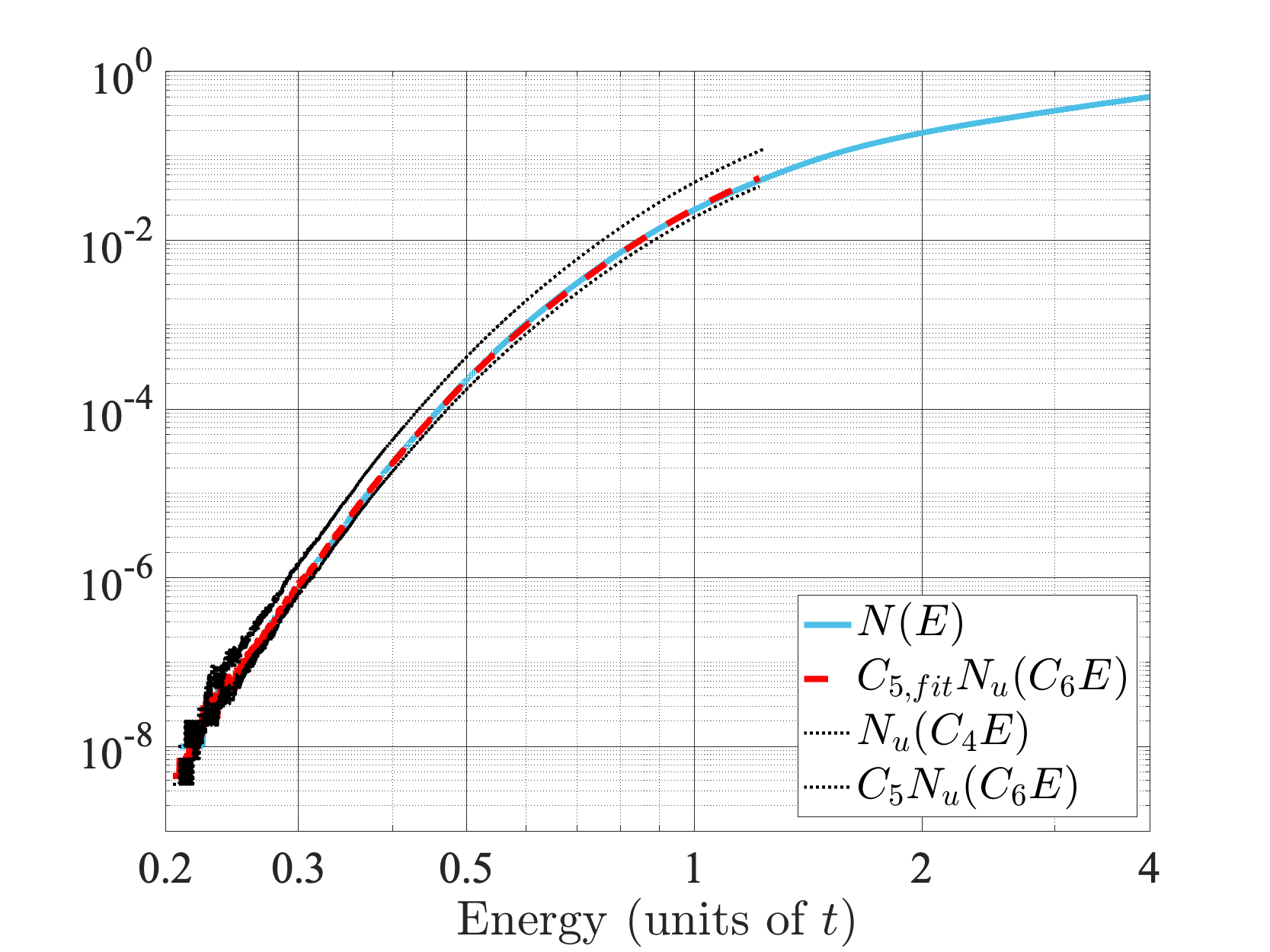}
\\(d)\\
\caption{(a) $N(E)$ (blue) averaged over 1000~random realizations, and averaged landscape law $N_u(E)$ (thick upper red), for a one-dimensional uniform Anderson tight-binding model of size $N=10^5$ and $V_{\max}=4$. (b) Standard deviation of the distribution of values of $\ln(N_u(CE)/N(E))$ for various values of $C$. The minimum around $C=0.82\approx 1/1.22$ provides the value of $C_6$. (c) Plot of $N_u(C_6 E)/N(E)$. The maximum in the noiseless part of the graph shows that one can take $C_5=1/2.77$. A best fit for $N(E)$ is obtained by taking the average value $C_{5,\textrm{fit} }\approx 1/2.23$. (d) Final comparison between the original $N(E)$ (blue), the best fit $C_{5,\textrm{fit}} N_u(C_6 E)$ (dashed red), and the two bounds from above and below $N_u(C_4E)$ and $C_5N_u(C_6E)$ (dotted lines).}
\label{fig:1D_uniform_average_4}
\end{figure}

To check the validity of our approach, we have investigated the role of the domain size for thesethese 1D
Hamiltonians (we could not run such a study in 2D because the computation time did not allow us to explore a large enough range of domain sizes). Domain sizes $N=10^3$, $10^4$, and $10^5$ were simulated, for both Anderson binary and Anderson uniform models. We also tested several values of the potential amplitude, $V_{\max}=1$, 2, and 4. For instance, Fig.~\ref{fig:1D_uniform_average_4} displays the analysis of the Anderson uniform model for $V_{\max}=4$. Once again, one can observe that the fit is excellent throughout the spectrum, justifying looking for constants that satisfy Eq.~\eqref{eq:landscape_fit} {(see Table~\ref{tab:constants_1} for the summary of these results)}.

\subsection{2D systems: results}

We then turned our study of 2D systems. The considered domain is a square of side length $L=1500$ which corresponds to $N=2.25 \times 10^6$ sites. Given that this system size is more than 10 times the size of the studied 1D systems, we could average only over 100~realizations for computational reasons. The fact that the side length of the system has been reduced by three orders of magnitude when going from 1D to 2D shifted considerably the lower bound of the energy range that could be explored. In the following simulations, we were unable to go below $E_{\min} \approx 0.2$.

\begin{figure}[h!]
\includegraphics[width=0.46\textwidth]{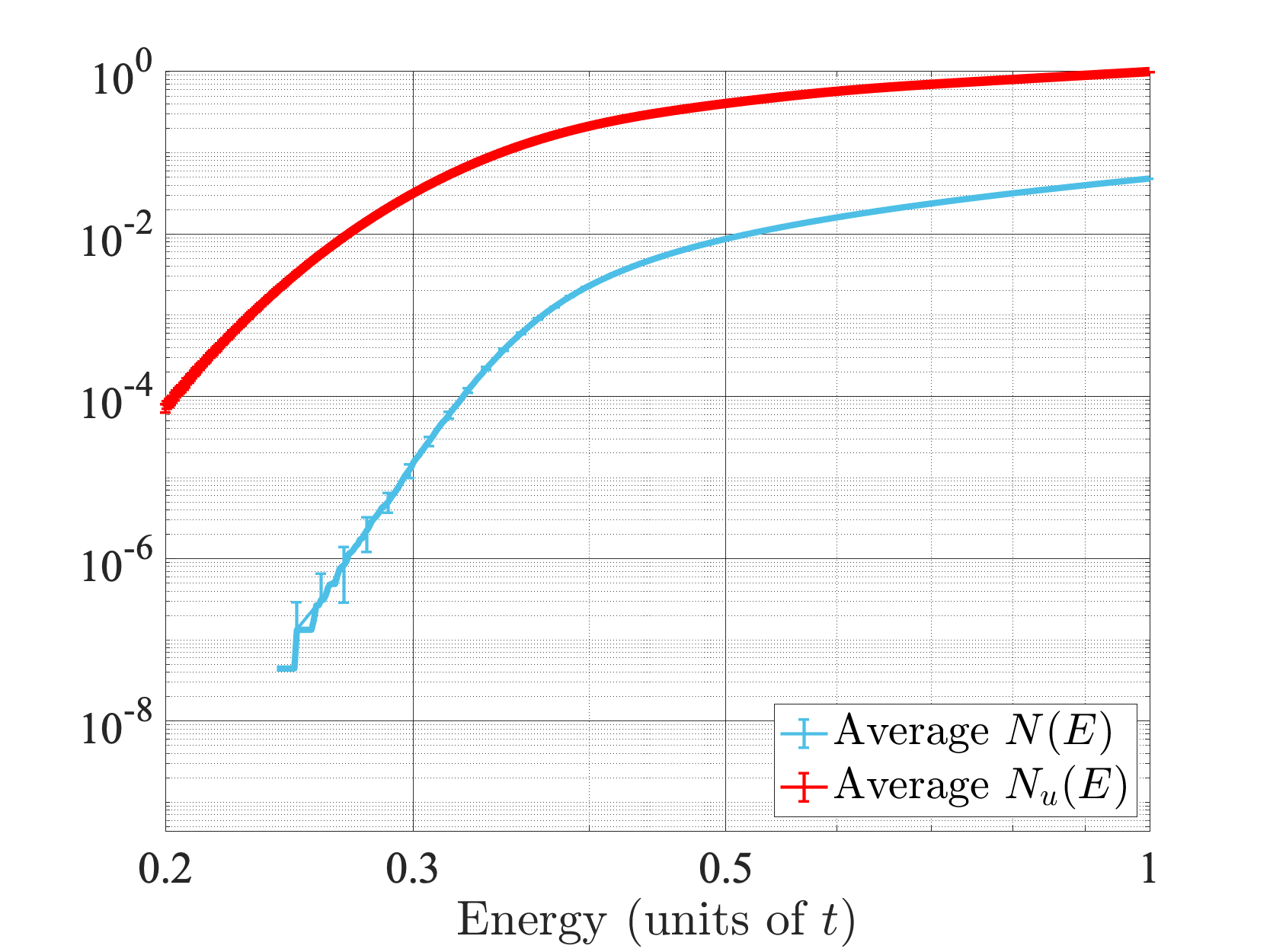}
\\(a)\\
\includegraphics[width=0.23\textwidth]{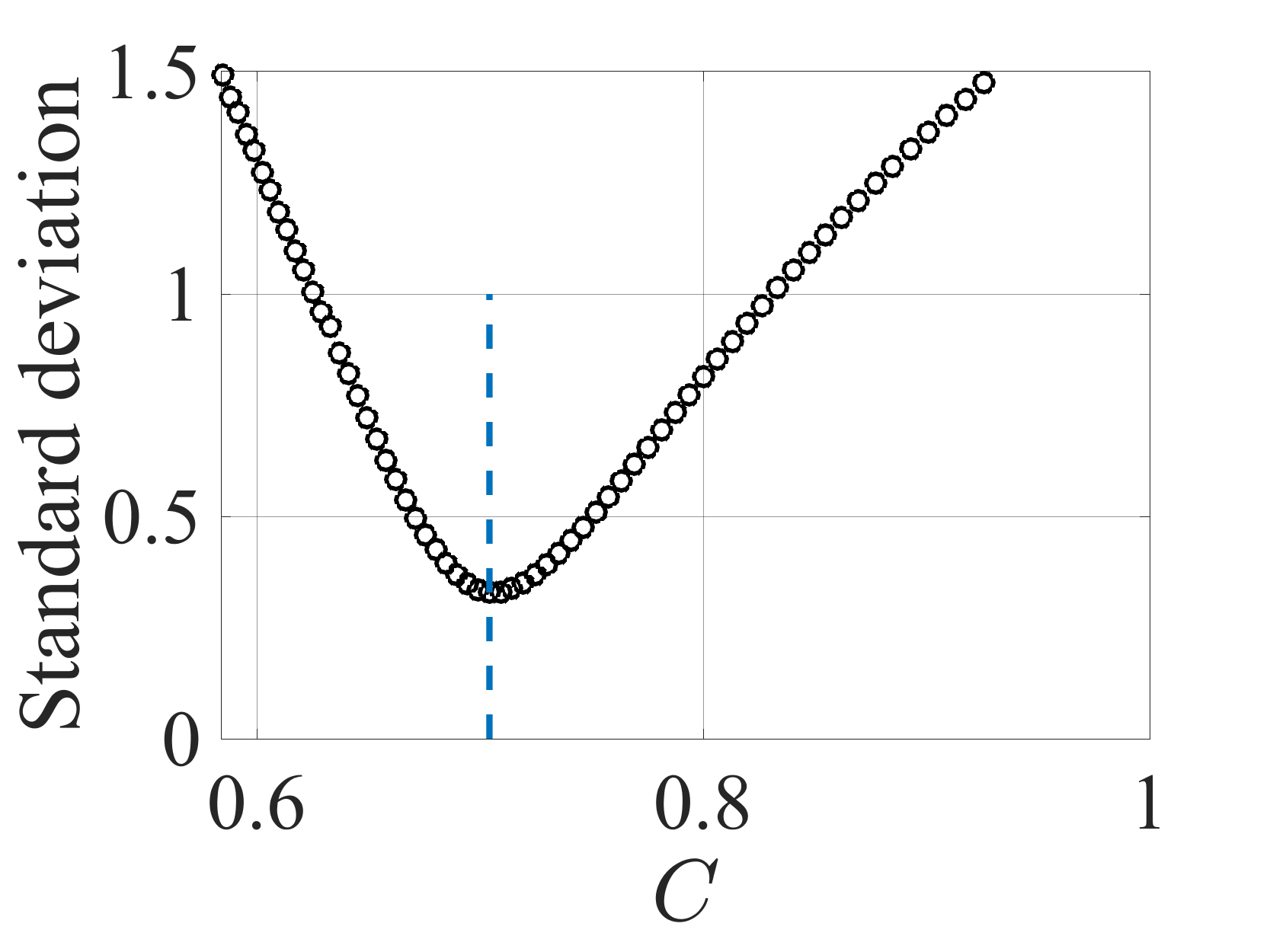}
\includegraphics[width=0.23\textwidth]{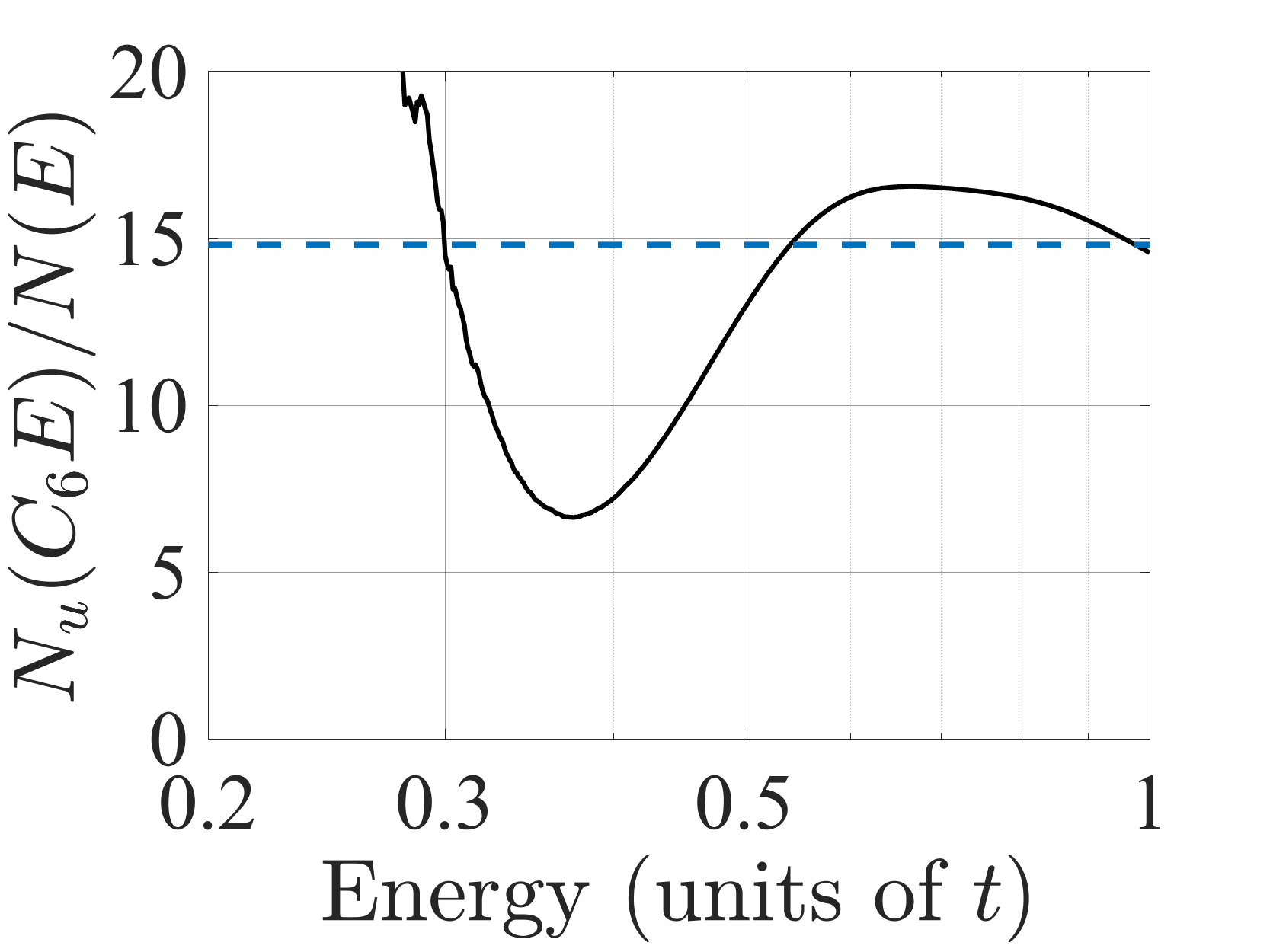}
\\ \hskip 3mm (b) \hskip 36mm (c)\\
\includegraphics[width=0.46\textwidth]{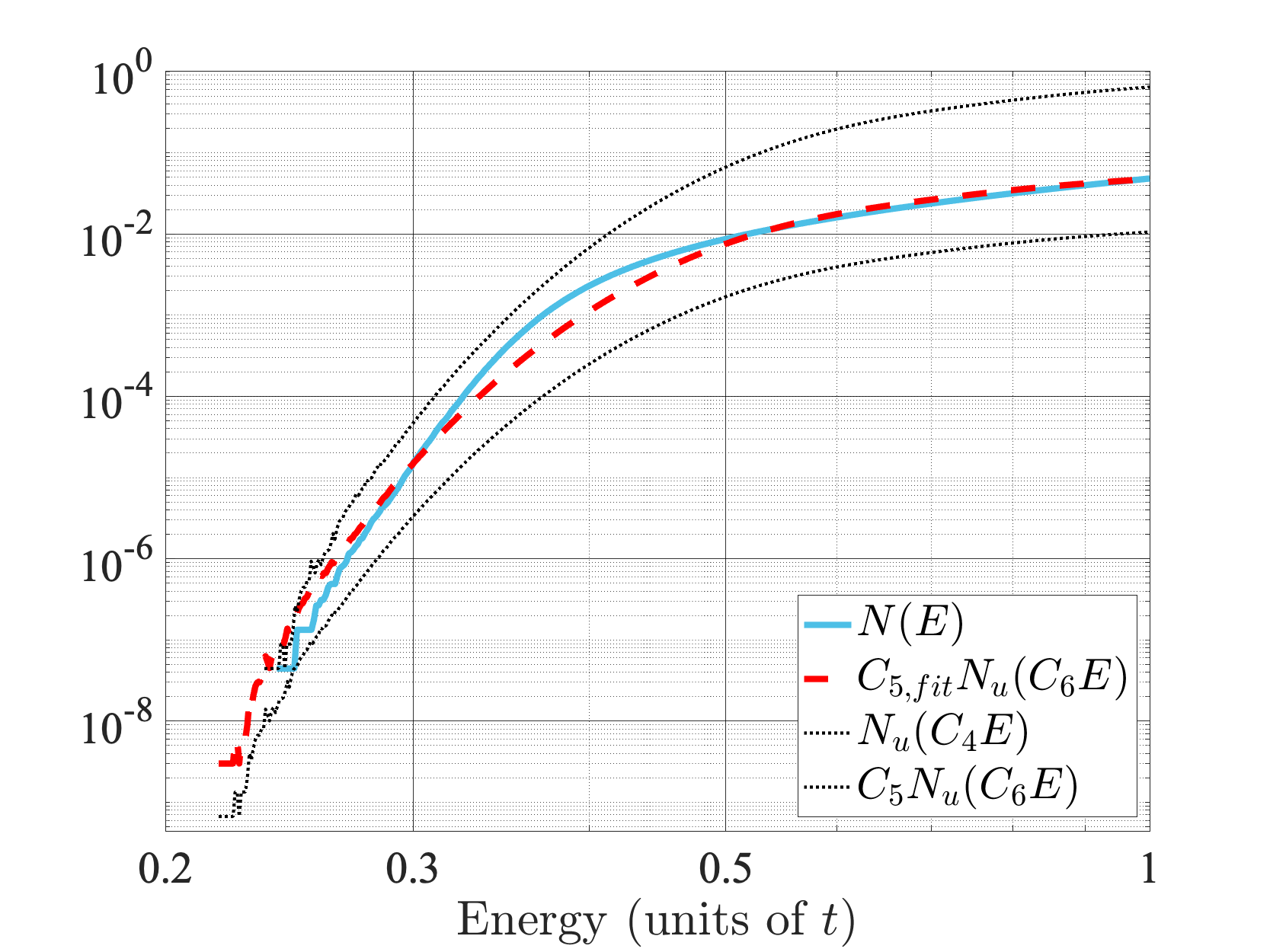}
\\(d)\\
\caption{(a)~$N(E)$ (blue) averaged over 100~random realizations, and averaged landscape law $N_u(E)$ (thick upper red), for a two-dimensional binary Anderson tight-binding model of size $N=(1500)^2$. (b)~Standard deviation of the distribution of values of $\ln(N_u(CE)/N(E))$ for various values of $C$. The minimum around $C=0.7\approx 1/1.42$ provides the value of $C_6$. (c)~Plot of $N_u(C_6 E)/N(E)$. Its maximum for $E>0.3$ shows that one can take $C_5=1/66.4$ which is also almost the best fit $C_{5,\textrm{fit} }\approx 1/14.8$. (d)~Final comparison between the original $N(E)$ (blue), the best fit $C_{5,\textrm{fit }} N_u(C_6 E)$ (dashed red), and the two bounds from above and below $N_u(C_4E)$ and $C_5N_u(C_6E)$ (dotted lines).}
\label{fig:2D_binary_average}
\end{figure}

\begin{figure}[h!]
\includegraphics[width=0.46\textwidth]{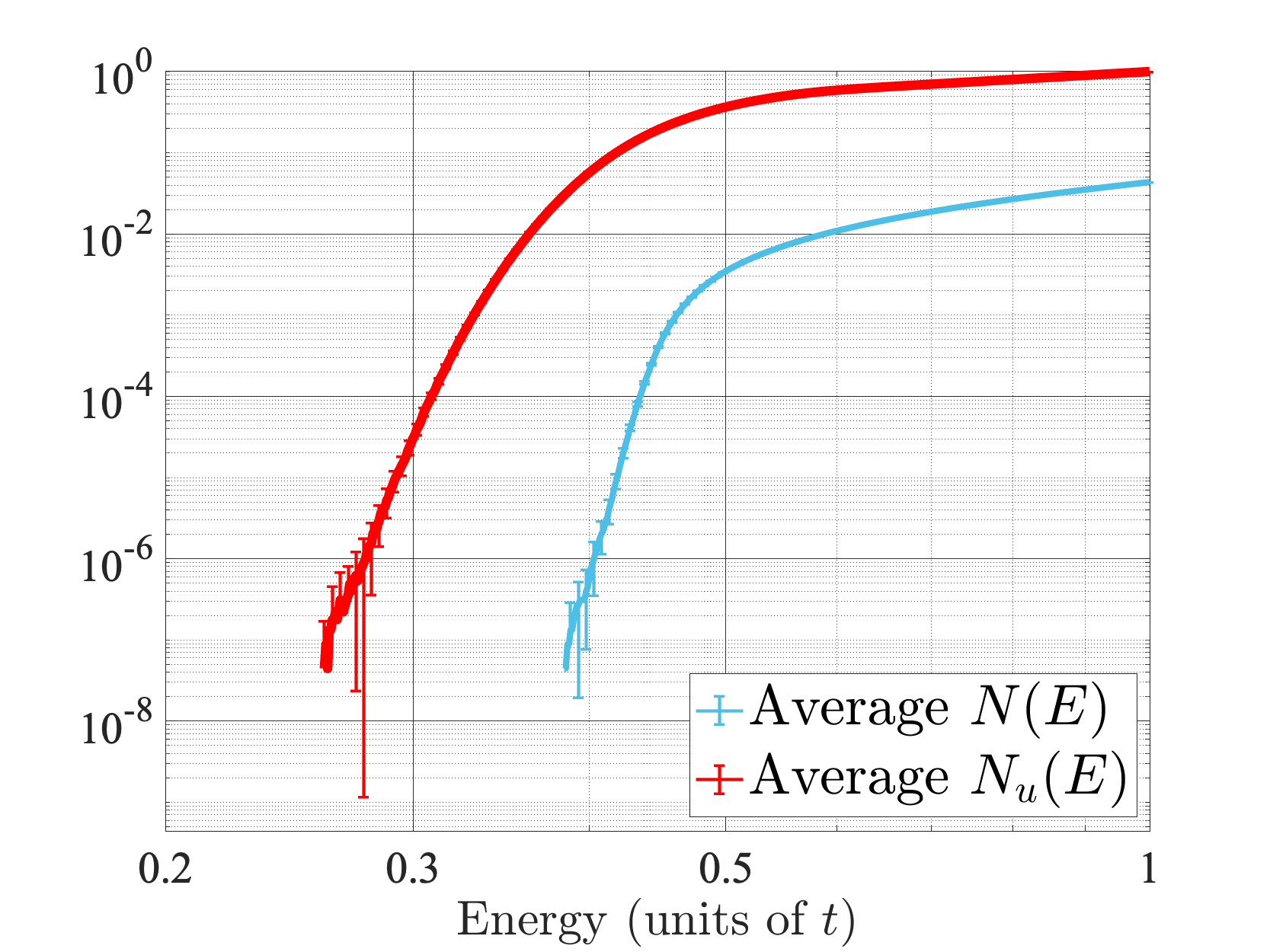}
\\(a)\\
\includegraphics[width=0.23\textwidth]{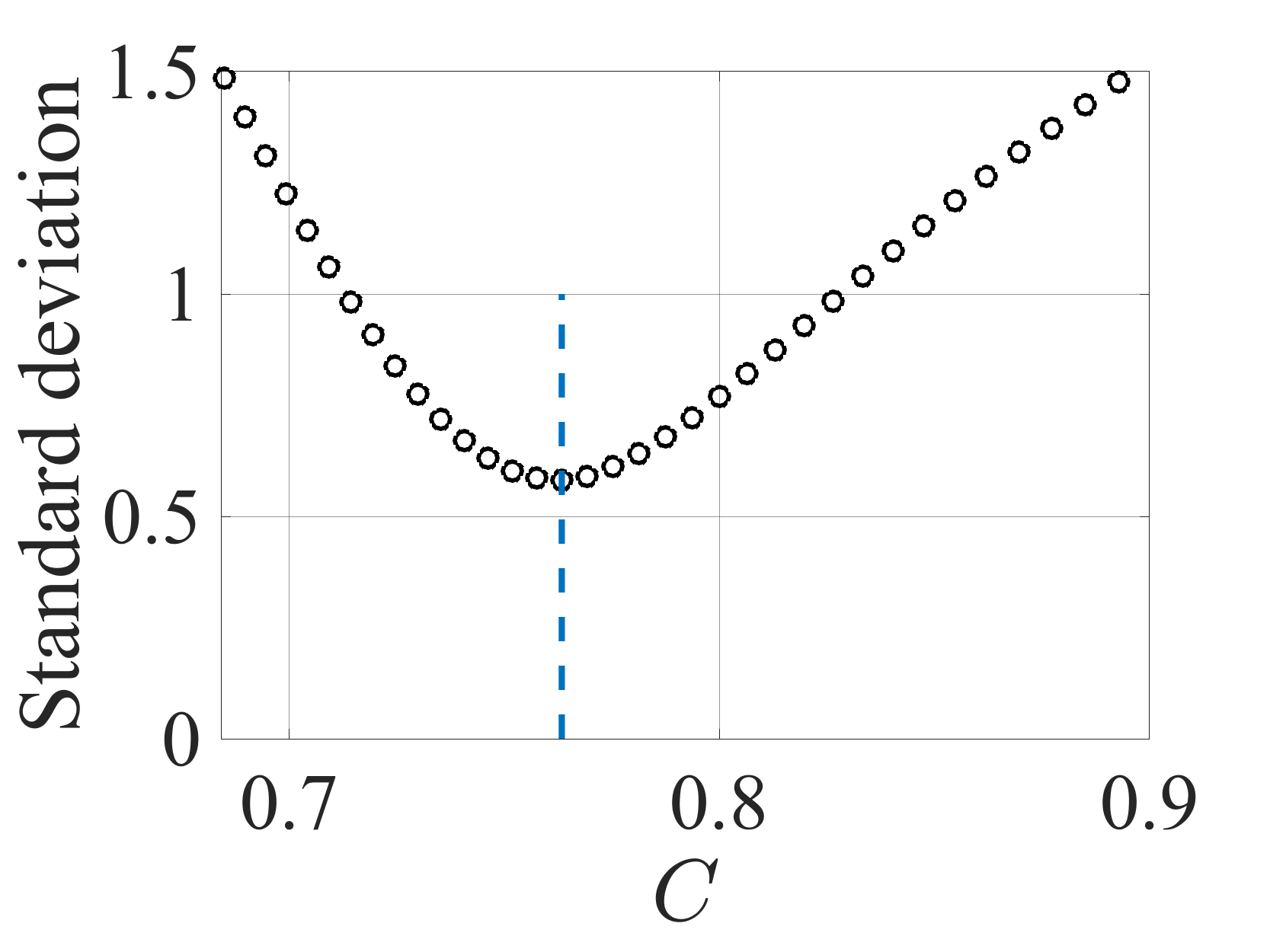}
\includegraphics[width=0.23\textwidth]{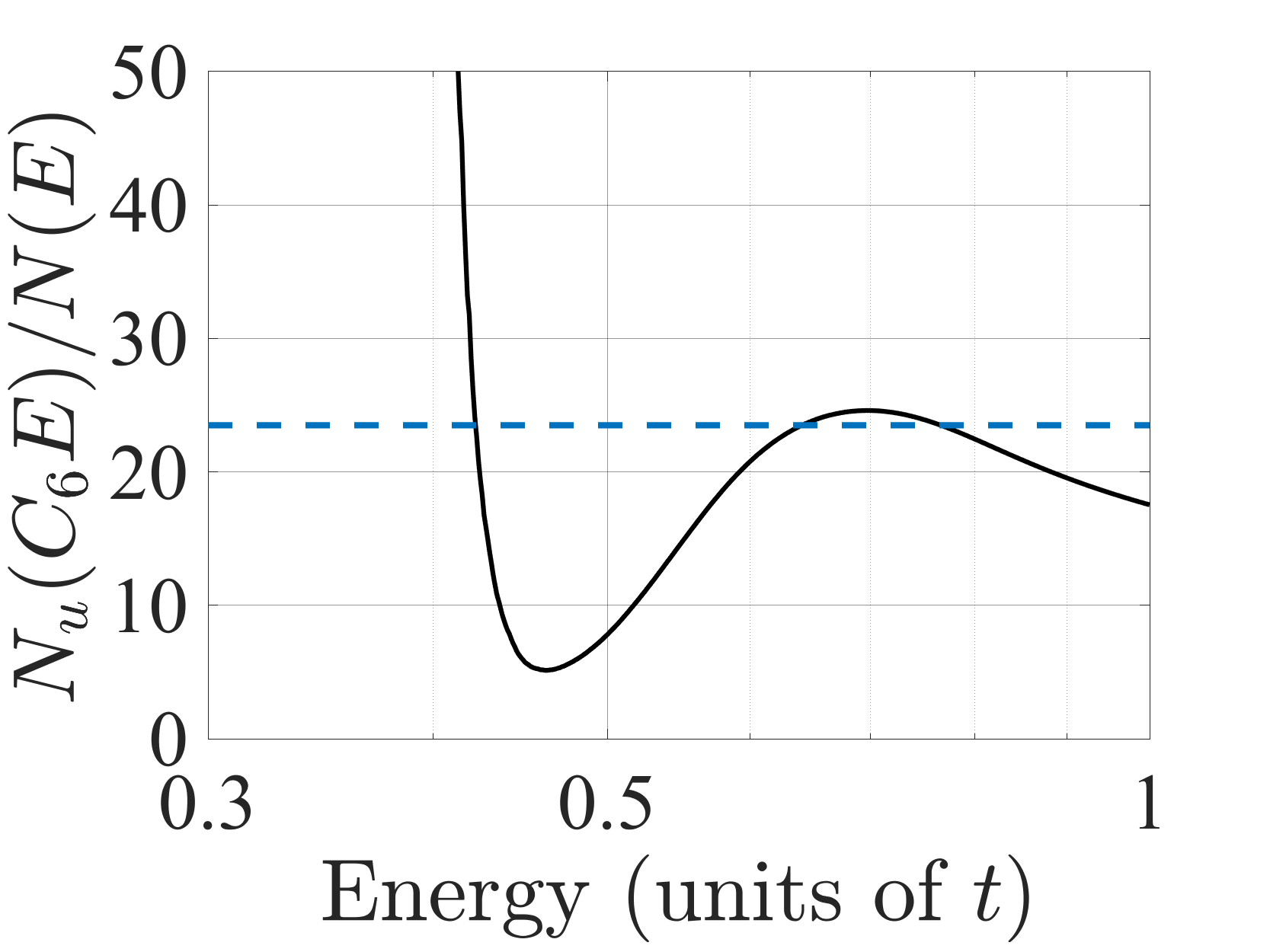}
\\ \hskip 3mm (b) \hskip 36mm (c)\\
\includegraphics[width=0.46\textwidth]{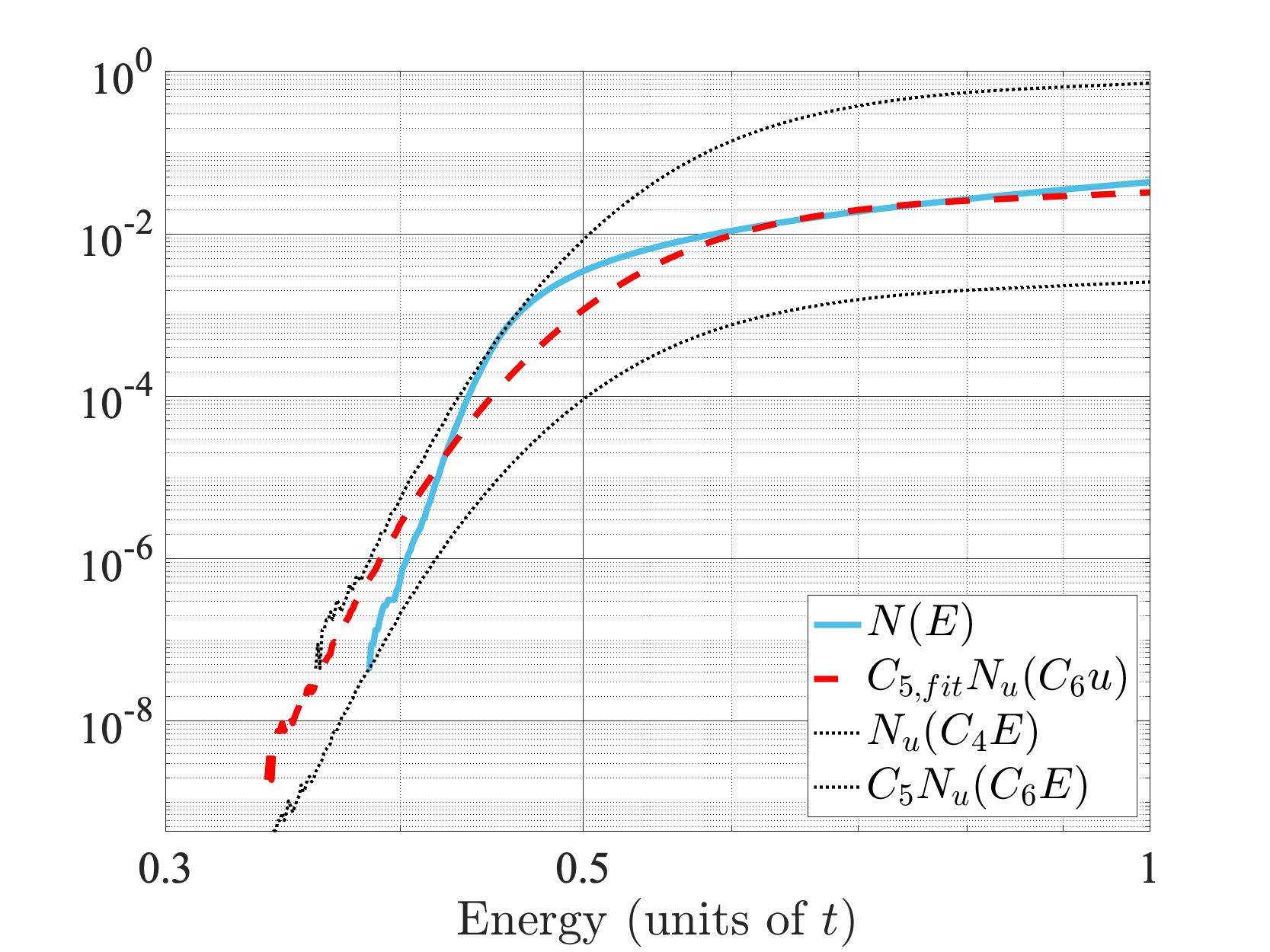}
\\(d)\\
\caption{(a) $N(E)$ (blue) averaged over 100~random realizations, and averaged landscape law $N_u(E)$ (thick upper red), for a two-dimensional uniform Anderson tight-binding model of size $N=(1500)^2$. (b) Standard deviation of the distribution of values of $\ln(N_u(CE)/N(E))$ for various values of $C$. The minimum around $C=0.76\approx 1/1.31$ provides the value of $C_6$. (c) Plot of $N_u(C_6 E)/N(E)$. Its maximum for $E>0.4$ shows that one can take $C_5=1/298$ while the best fit leads to $C_{5,\textrm{fit} }\approx 1/23.5$. (d) Final comparison between the original $N(E)$ (blue), the best fit $C_{5,\textrm{fit }} N_u(C_6 E)$ (dashed red), and the two bounds from above and below $N_u(C_4E)$ and $C_5N_u(C_6E)$ (dotted lines).}
\label{fig:2D_uniform_average}
\end{figure}

Figure~\ref{fig:2D_binary_average} displays the analysis for a 2D binary Anderson model. The constants extracted from the analysis are $C_4 = 1/1.53$, $C_{5,\textrm{fit} } = 1/14.5$, $C_6 = 1/1.42$. The agreement between $N(E)$ and the rescaled landscape law is still good in the whole energy range, even though one can see now that the ratio $N_u(C_6 E)/N(E)$ oscillates much more than in the 1D case [see Fig.~\ref{fig:1D_binary_average}(c)]. This observation is even more marked in the case of the 2D uniform Anderson case (see Fig.~\ref{fig:2D_uniform_average}). One can observe that the upper and lower bounds are significantly apart in this case, especially at larger energy. This reflects the fact that the prefactor of $N_u(E)$ has to be different at low and high energy to approximate $N(E)$ accurately.

\section{Analysis}

Table \ref{tab:constants_1} summarizes the values obtained for the constants in all cases. For the three 1D cases displayed in Figs~\ref{fig:1D_binary_average}-\ref{fig:1D_uniform_average_4}, we performed an error bar analysis by splitting the 1000~realizations into 20 samples of 50 realizations, and computing the constants separately for each sample. The numbers to the right of the symbol~$\pm$ correspond to twice the standard deviation.

This table triggers several comments. First, the values of the constants $C_4$, $C_5$, $C_{5,\textrm{fit}}$, $C_6$ do not seem to depend at all on the domain size (for the same potential law). For instance, for $V_{\max}=1$ and a binary disorder, the values of $C_4^{-1}$ are 1.26, 1.27, 1.26, for $L=10^5$, $10^4$, $10^3$, respectively, while for the same $V_{\max}$ and a uniform disorder the corresponding values are 1.28, 1.28, and 1.27. Similarly, the values of $C_6^{-1}$ are 1.11, 1.1, 1.08 for the same domain sizes for a binary disorder and 1.19, 1.18, and 1.16 for a uniform disorder. The values of $C_{5,\textrm{fit}}^{-1}$ are 4.08, 4.15, 4.27 for a binary disorder, and 3.94, 4.05, 4.29 for a uniform disorder. Finally, the values of $C_5^{-1}$ are 5.45, 8.18, 5.96 for the binary disorder and 4.85, 7.81, 8.86 for a uniform disorder. These last values are slightly more dispersed, and the reason is that they are determined by the maxima of the curves in frames~(c) (Figs.~\ref{fig:1D_binary_average}--\ref{fig:2D_uniform_average}) which depend on the accuracy of the data in the lower part of the spectrum. This dispersion justifies looking for the values of $C_{5,\textrm{fit}}$ which are much more reliable than the ones of $C_5$.

Second, the values of $1/C_4$ are quite close to the value $1 + d/4$, where $d$ is the ambient dimension. This value arises in the localization landscape approach as the ratio between a local fundamental eigenvalue inside a localization region and the local minimum of the effective potential $W=1/u$~\cite{Arnold2019, Arnold2019b}. From the definition of $N_u(E)$, at a given energy $E$, a $d$-cube of side length $E^{-1/2}$ contributes to $N_u(E)$ only if $\min(W)$ inside the cube is smaller than $E$. In that situation, one would expect a local fundamental eigenvalue roughly at $(1+d/4) W_{\min}$. Consequently, there is a natural multiplicative shift in energy between $N(E)$ and $N_u(E)$, by a factor $1+d/4$. This is what is found in our 1D and 2D simulations. One has to note that this shift has already been observed in a very different model, namely, the pieces model in which a 1D system is partitioned into sub-intervals of random length following a Poisson law~\cite{Comtet2020, Filoche2020}.

Third, the values of $C_6$ follow rather closely those of $C_4$, being only slightly larger ($C_6^{-1} < C_4^{-1}$). We observe that the values of $C_6$ are closer to those of $C_4$ for Anderson uniform models.

Fourth, the results displayed in Table~\ref{tab:constants_1} help us understand the influence of the disorder strength~$V_{\max}$. To that end, we have set $V_{\max}=1$, 2, or 4 for Anderson binary and Anderson uniform models in 1D and 2D. The theory developed in Ref.~\cite{David2019landscape} states that the constants involved in the bounds and which depend on the potential should, in fact, not depend of its maximum value but rather on its average value. However, in both Anderson binary and Anderson uniform models, the average value of the potential is $\ev{V} = V_{\max}/2$, so it is still directly determined by the disorder strength. In all computed cases, we observe that the values of $C_4$ and $C_6$ remain almost unchanged while the value of $C_{5,\textrm{fit}}$ appears to be roughly proportional to $V_{\max}^{1/2}$, which is a natural scaling in the problem at hand.

\begin{table*}[t]
\centering
\begin{tabular}{@{}cccccccc@{}}
    \toprule[1pt]
    \multicolumn{2}{c}{Model} & $L$ & $V_{\max}$ & $1 / C_4$ &  $1/ C_5$ & $1 / C_{5,\textrm{fit}}$ & $1 / C_6$ \\
    \midrule[1pt]
    \multirow{10}{*}{1D} & \multirow{5}{*}{binary} & \multirow{3}{*}{$10^5$} & 1 & 1.26$\,\pm0.05$ & 5.45\,$\,\pm\,5$ & 4.08\,$\,\pm\,0.15$ & 1.11\,$\,\pm\,0.02$\\
    \cmidrule[1pt]{4-8}
    & & & 2 & 1.3 & 3.78 & 3.03 & 1.2 \\
    \cmidrule[1pt]{4-8}
    & & & 4 & 1.26 & 2.91 & 2.04 & 1.32 \\
    \cmidrule[1pt]{3-8}
    & & $10^4$ & \multirow{2}{*}{1} & 1.27 & 8.18 & 4.15 & 1.1 \\
    \cmidrule[1pt]{3-3} \cmidrule[1pt]{5-8}
    & & $10^3$ & & 1.26 & 5.96 & 4.27 & 1.08 \\
    \cmidrule[1pt]{2-8}
    &\multirow{5}{*}{uniform} & \multirow{3}{*}{$10^5$} & 1 & 1.28$\,\pm0.02$ & 4.85$\,\pm4.3$ & 3.94$\,\pm0.2$ & 1.19$\,\pm0.02$ \\
    \cmidrule[1pt]{4-8}
    & & & 2 & 1.24 & 8.14 & 3.36 & 1.19 \\
    \cmidrule[1pt]{4-8}
    & & & 4 & 1.24$\,\pm0.03$ & 2.77$\,\pm1.2$ & 2.23$\,\pm0.05$ & 1.22$\,\pm0.01$ \\
    \cmidrule[1pt]{3-8}
    & & $10^4$ & \multirow{2}{*}{1} & 1.28 & 7.81 & 4.05 & 1.18 \\
    \cmidrule[1pt]{3-3} \cmidrule[1pt]{5-8}
    & & $10^3$ & & 1.27 & 8.86 & 4.29 & 1.16 \\
   	\midrule[2pt]
    \multirow{4}{*}{2D} & \multirow{2}{*}{binary} & \multirow{4}{*}{1500} & 1 & 1.53 & 66.4 & 14.8 & 1.42 \\
    \cmidrule[1pt]{4-8}
    & & & 2 & 1.54 & 33.8 & 9.00 & 1.44 \\
    \cmidrule[1pt]{2-2} \cmidrule[1pt]{4-8}
    & \multirow{2}{*}{uniform} & & 1 & 1.39 & 298 & 23.5 & 1.31 \\
    \cmidrule[1pt]{4-8}
    & & & 2 & 1.47 & 4.64 & 1.83 & 1.48 \\
    \bottomrule[1pt]
\end{tabular}
    \caption{Summary of the values found for the constants $C_4$, $C_5$, $C_{5,\textrm{fit}}$, and $C_6$. The Anderson models are one- or two-dimensional, with binary or uniform random laws. $L$ is the side length (so the system size is $|\Omega|=L^d$), and $V_{\max}$ is the disorder strength. For the three cases presented in Figs.~\ref{fig:1D_binary_average}--\ref{fig:1D_uniform_average_4}, error bars were computed. The numbers displayed after the $\pm$ symbol correspond to two standard deviations. The 2D computations performed were with too few realizations to derive meaningful statistics.}
    \label{tab:constants_1}
\end{table*}

\subsection{Scaling analysis near $E=0$}

One can mention an alternative way of relating $N(E)$ and $N_u(E)$ at the bottom of the spectrum, and therefore of extracting the constants involved in~Eq.~\eqref{eq:landscape_fit}. This consists of using a corollary of the Landscape Law pertaining to the scaling at low energies for all random i.i.d. potentials. In Ref.~\cite{David2019landscape}, it is shown that, in the case of an Anderson tight-binding model where the on-site potential values $\{V_i\}$ follow a random law of cumulative distribution function~$F$ (i.e., the probability to have $V_i \leq E$ is $F(E)$), there exist constants $\gamma_1$, $\gamma_2$, $\gamma_3$, $\gamma_4$, $c_1$, $c_2$ such that
\begin{equation}\label{eq:scaling}
\gamma_3 F\left(c_2 E \right)^{\gamma_4 E^{-\frac{d}{2}}} \leq \frac{N(E)}{E^{\frac{d}{2}}} \leq \gamma_1 F\left(c_1 E \right)^{\gamma_2 E^{-\frac{d}{2}}}\,.
\end{equation}
More specifically, in the case of binary and uniform random laws, one has $F(E) = 1/2$ and $F(E)=E$ for $0<E<1$, respectively. Therefore, with a slight change of the meaning of $\gamma_4$,
\begin{align}\label{eq:log_scaling_binary}
\gamma_3 \, e^{\gamma_4 E^{-\frac{d}{2}}} &\leq N(E) \, E^{-\frac{d}{2}} \leq \gamma_1 \, e^{\gamma_2 E^{-\frac{d}{2}}} \,,\\
\gamma_3 e^{\gamma_4 E^{-\frac{d}{2}} |\ln(E)|} &\leq N(E) \, E^{-\frac{d}{2}} \leq \gamma_1 e^{\gamma_2 E^{-\frac{d}{2}} |\ln(E)|} \,.
\end{align}

Let us consider the binary Anderson model. The inequality~\eqref{eq:log_scaling_binary} can be rewritten as
\begin{align}\label{eq:scaling_binary}
\gamma_4 + \ln(\gamma_3) E^{\frac{d}{2}} &\leq E^{\frac{d}{2}} \ln\left(N(E) \, E^{-\frac{d}{2}}\right) \nonumber\\
&\leq \gamma_2 + \ln(\gamma_1) E^{\frac{d}{2}} \,.
\end{align}
In other words, the quantity $E^{\frac{d}{2}} \ln\left(N(E) \, E^{-\frac{d}{2}}\right)$ can be bounded between two affine functions of $E^{\frac{d}{2}}$. In the uniform Anderson model, a similar expression holds with a logarithmic correction:
\begin{align}\label{eq:scaling_uniform}
\gamma_4 + \ln(\gamma_3) \frac{E^{\frac{d}{2}}}{|\ln(E)|} &\leq \frac{E^{\frac{d}{2}}}{|\ln(E)|} \ln\left(N(E) \, E^{-\frac{d}{2}}\right) \nonumber\\
 &\leq \gamma_2 + \ln(\gamma_1) \frac{E^{\frac{d}{2}}}{|\ln(E)|} \,.
\end{align}

\begin{figure}[ht!]
\includegraphics[width=0.3\textwidth]{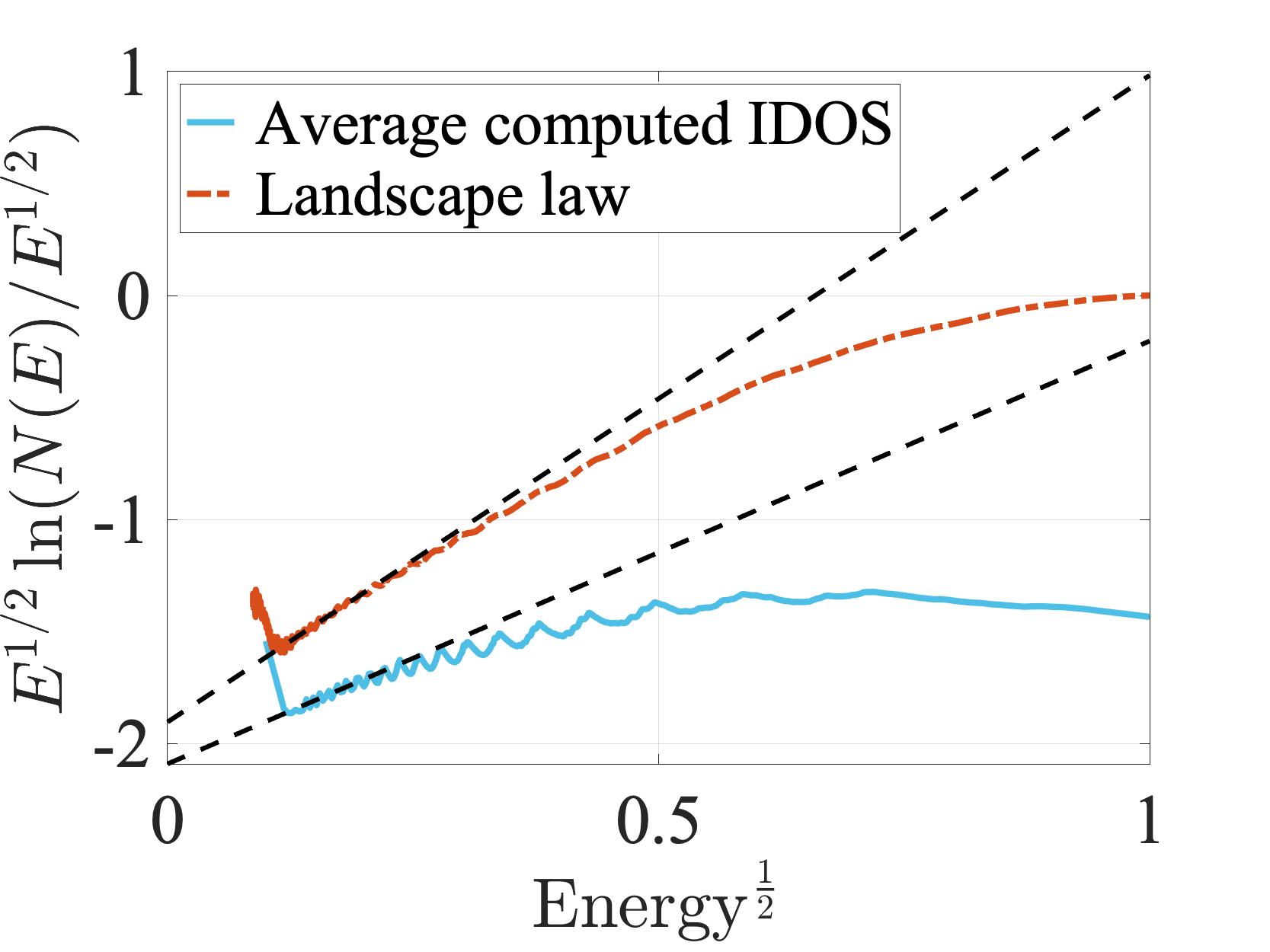}
\includegraphics[width=0.3\textwidth]{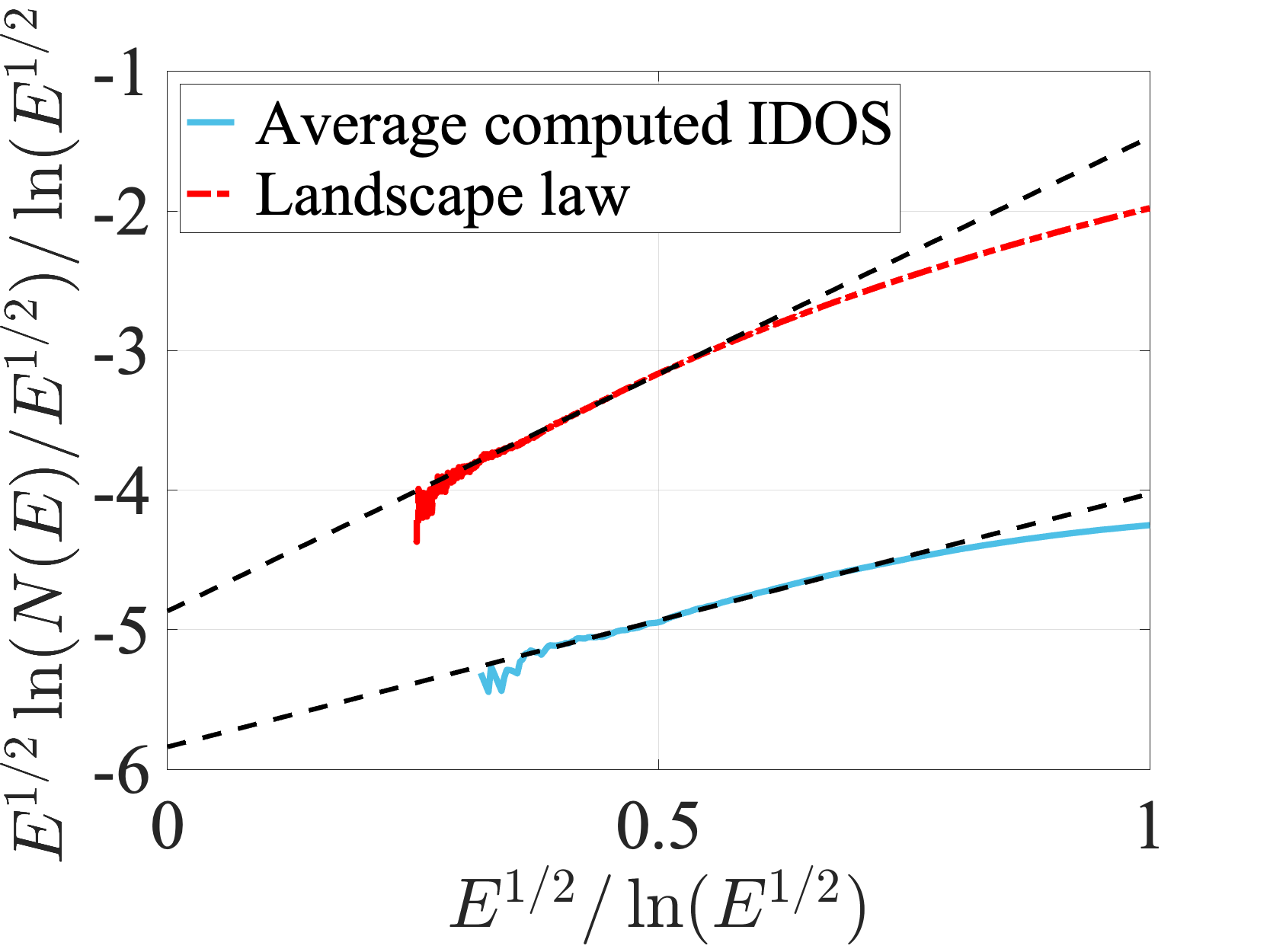}
%\\ \hskip 3mm (a) \hskip 37mm (b)\\
\includegraphics[width=0.3\textwidth]{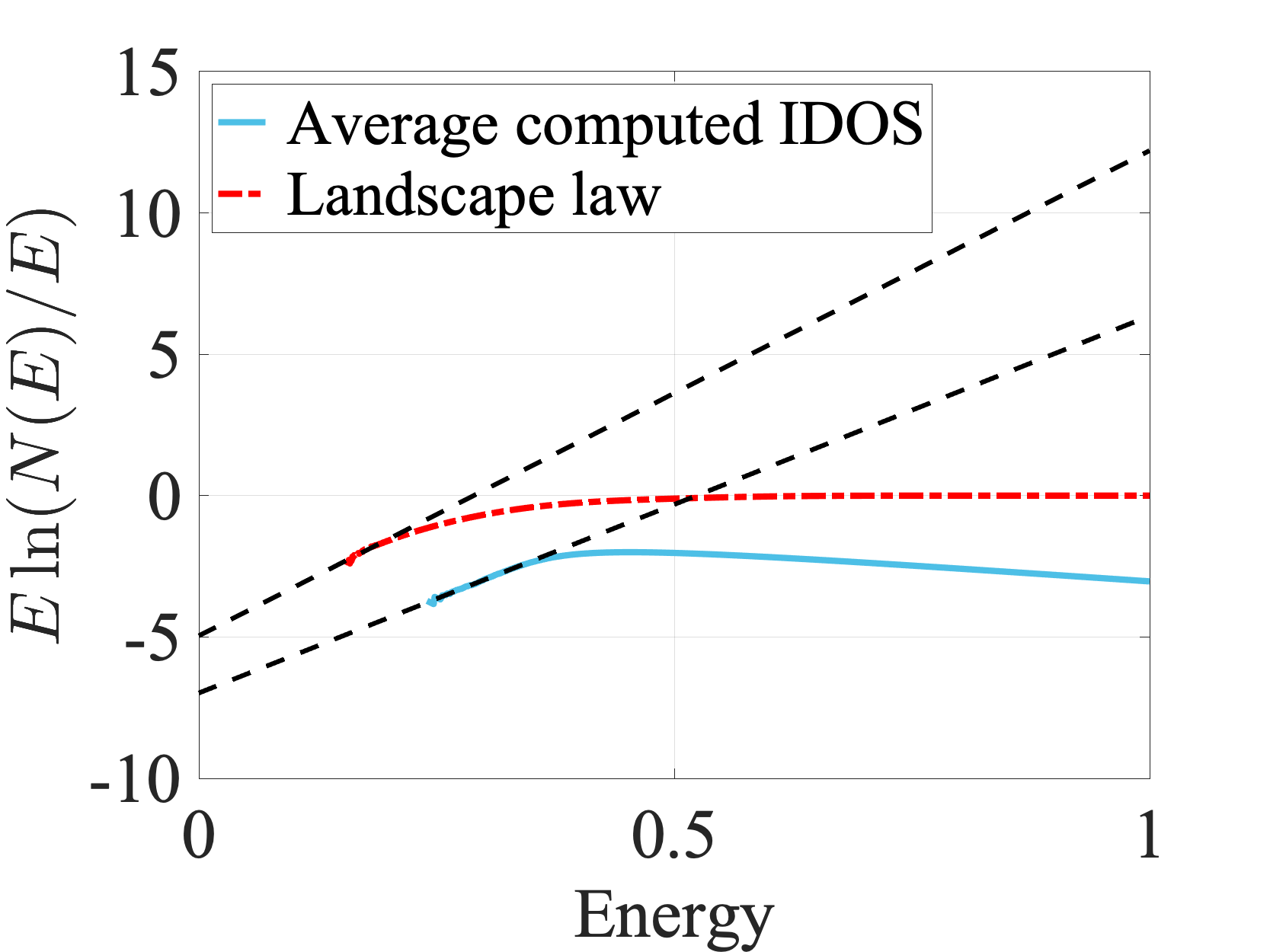}
%\\ \hskip 3mm (c) \\
\caption{Scaling behavior of $N(E)$ (blue line) and $N_u(E)$ (red dashed line) near $E=0$. The straight lines correspond to the asymptotic linear behaviors appearing in Eqs.~\eqref{eq:scaling_binary} and \eqref{eq:scaling_uniform}. (a) 1D binary Anderson model. (b) 1D uniform Anderson model. (c) 2D binary Anderson model. All quantities are dimensionless.}
\label{fig:scaling_analysis}
\end{figure}

Figure~\ref{fig:scaling_analysis} displays the graphs of these quantities near $E=0$ in three different cases already examined: (i) a 1D binary Anderson model (cf. Fig.~\ref{fig:1D_binary_average}), (ii) a 1D uniform Anderson model (cf. Fig.~\ref{fig:1D_uniform_average}), and (iii) a 2D uniform Anderson model (cf. Fig.~\ref{fig:2D_uniform_average}). In each case, the values of $\gamma_1$, $\gamma_2$, $\gamma_3$, $\gamma_4$ are extracted from the scaling behavior of $N(E)$ and $N_u(E)$ (the linear scaling relations expressed in Eqs.~\eqref{eq:scaling_binary} and \eqref{eq:scaling_uniform} are shown in dotted lines in the graphs), and then used to compute the effective values of $C_5$ and $C_6$ relating $N$ to $N_u$. The findings are grouped in Table~\ref{tab:constants_2}. One has to underline that the huge computation time required to reach very low values of $E$ limited the accuracy and the range of the data on which the scaling behavior could efficiently be tested and led to significant error bars. It precluded us from performing this analysis in the 2D uniform Anderson model. Even in the 2D binary Anderson model [Fig.~\ref{fig:scaling_analysis}(c)], the scaling is observed for a very limited range of energies. Yet, we observe in 1D the scaling predicted by the mathematical proof in Ref/~\cite{David2019landscape} [(see Fig.~\ref{fig:scaling_analysis}(a) and \ref{fig:scaling_analysis}(b)]. The parameters $C_5$ and $C_6$ are consistent with the values reported in Table~\ref{tab:constants_1}, confirming that $N_u$ can be used through Eq.~\eqref{eq:landscape_fit} to approximate $N(E)$ throughout the spectrum. Finally, in 2D [Fig.~\ref{fig:scaling_analysis}(c)], the discrepancy in the values of $C_5$ clearly indicates that one cannot find a single prefactor satisfying Eq.~\eqref{eq:landscape_fit}, and that this prefactor, in fact, has to be modified into a very slowly varying function from $E=0$ to the largest eigenvalue.

\begin{table}[h!]
\centering
\begin{tabular}{@{}ccccccc@{}}
	\cmidrule{4-7}
	\multicolumn{3}{c}{} & \multicolumn{2}{c}{Table I} & \multicolumn{2}{c}{scaling} \\
    \midrule[1pt]
    \multicolumn{2}{c}{Model} & $V_{\max}$ & $1 / C_{5,\textrm{fit}}$ &  $1/ C_6$ & $1 / C_5$ & $1 / C_6$ \\
    \midrule[1pt]
    \multirow{2}{*}{1D} & binary & 1 & 4.08 & 1.11 & 2.55 & 1.2 \\
    \cmidrule[1pt]{2-7}
     & uniform & 1 & 3.95 & 1.19 & 4.05 & 1.44 \\
    \midrule[1pt]
     2D & binary & 1 & 14.5 & 1.42 & 0.84 & 1.62 \\
    \bottomrule[1pt]
\end{tabular}
    \caption{Comparisons between the constants $C_{5,\textrm{fit}}$ and $C_6$ from Table~\ref{tab:constants_1} and the constants $C_5$ and $C_6$ obtained from the scaling analysis near $E=0$.}
    \label{tab:constants_2}
\end{table}

\section{Conclusion}

\begin{figure}[ht!]
\vskip 3mm
\includegraphics[width=0.46\textwidth]{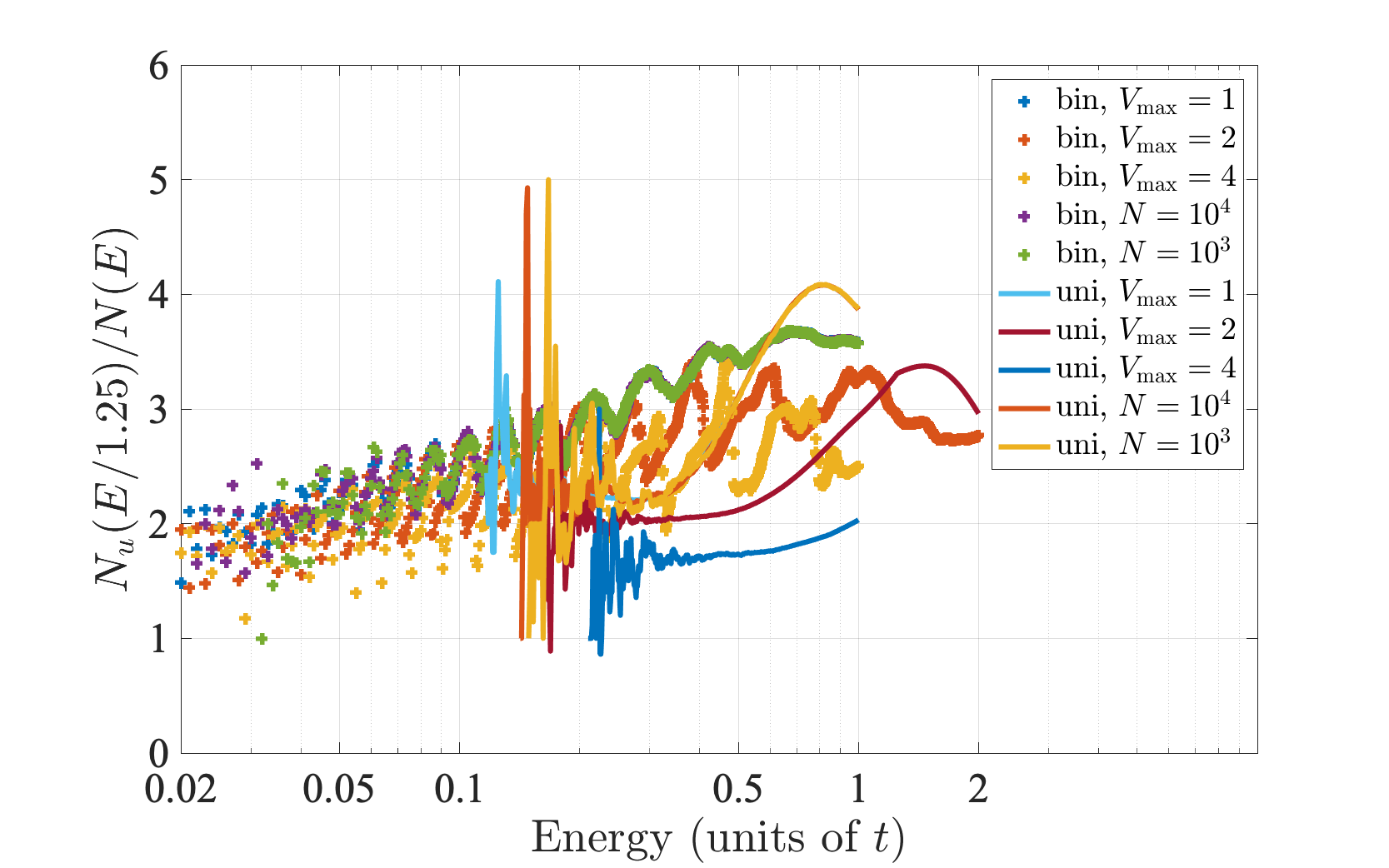}\\
\includegraphics[width=0.46\textwidth]{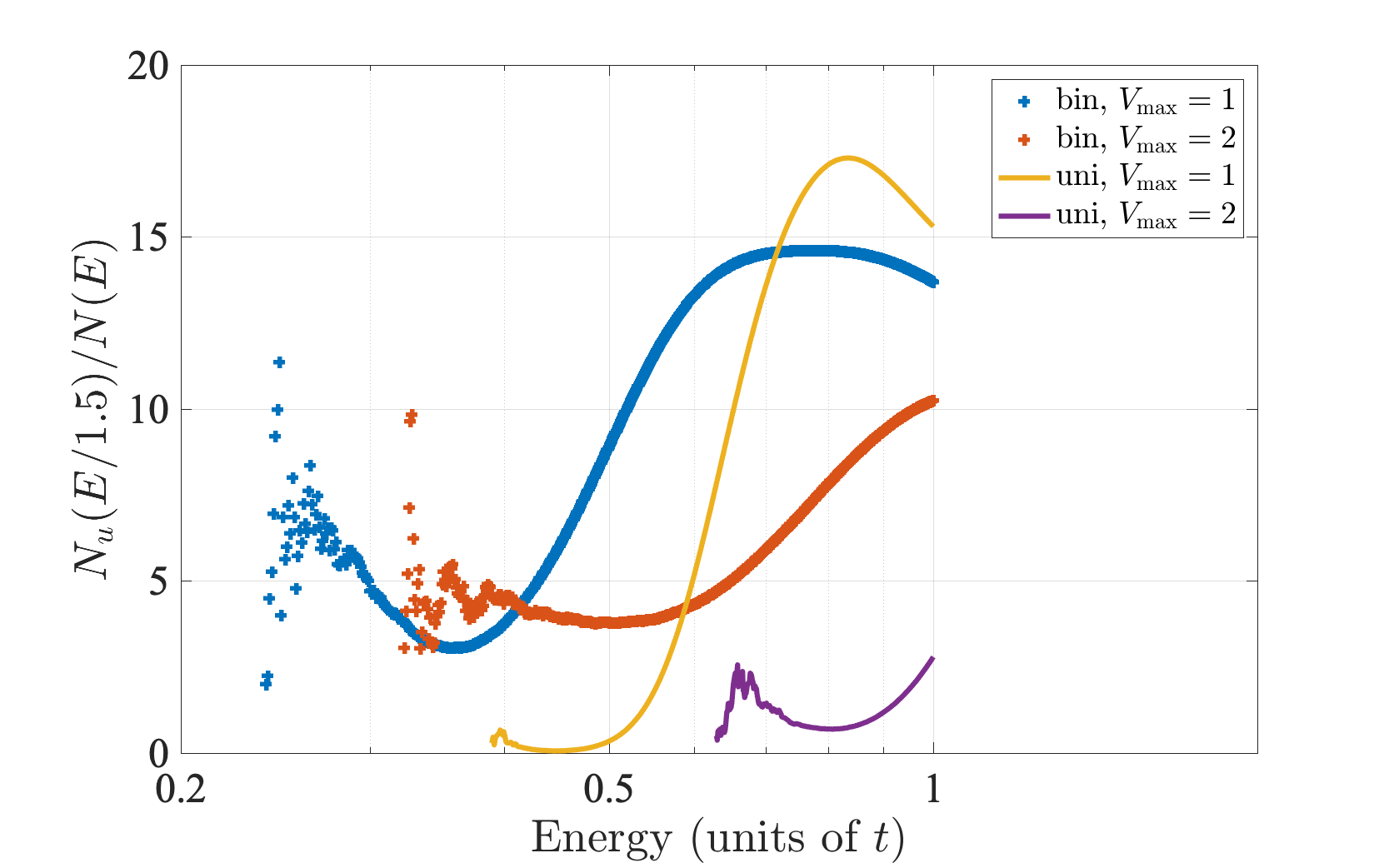}
\caption{Ratio $N_u\left( E/(1+d/4) \right)/N(E)$ plotted as a function of the energy $E$. Top: For all 1D models reported in Table~\ref{tab:constants_1}. Bottom: The same plot for all 2D models reported in Table~\ref{tab:constants_1}.}
\label{fig:universal_1}
\end{figure}

The general picture that emerges from this exhaustive numerical study is that $N_u(E)$ follows very closely the behavior of the actual IDOS $N(E)$ throughout the entire energy range while at the same time being much easier to compute and to handle. Although it is not always possible to approximate $N(E)$ through one single expression such as Eq.~\ref{eq:landscape_fit}, one can wonder whether we could obtain a very good estimate with almost universal constants. First, remember that the values found for the constant $C_6$ are all very close to $1/(1+d/4)$, for a reason expressed in Ref.~\cite{Arnold2019b}. A natural universal approximation for $N(E)$, without any fitting parameter, could thus be proportional to $N_u\left( E/(1+d/4) \right)$. To test this hypothesis, we plot the ratio $N(E)/N_u\left( E/(1+d/4) \right)$ vs $E$ for all cases reported in Table~\ref{tab:constants_1}, see Fig.~\ref{fig:universal_1}.

We observe that in 1D, all curves for all tested potentials (binary or Anderson models, different disorder strengths, different system sizes) follow the same pattern, i.e., a slow evolution from a value close to 2 at low energy to a value close to 4 at larger energy. In 2D, the structure is similar with a wider dynamics, from about 1 to about 16. This means that while the IDOS $N(E)$ spans several orders of magnitude (about 6 to 10 in our examples), the function $N_u(E/(1+d/4))$ always remains remarkably close to $N(E)$. The prefactor appears to be different in the low- and high-energy regimes, although it seems within reach, at least in 1D, to derive a very slowly varying function of the energy that would account for this change of prefactor. This change of prefactor between the low- and the high-energy regimes can be understood. One knows that, at least in the continuous setting, $N(E)/N_u(E)$ is equivalent to $\omega_d/(2\pi)^d$ at higher energy (so independent of the potential), with $\omega_d$ the volume of the unit ball in dimension~$d$. Therefore, as soon as $E > V_{\max}$, all cubes satisfy the condition in Eq.~\eqref{eq:landscape_law} and $N_u(E) = E^{d/2}$. On the other hand, in the low energy limit, one expects $N(E)$ to behave as $N_u(E/(1+d/4))$ which implies a different prefactor depending on the type of potential.

In conclusion, we have presented here a function called the landscape law which provides bounds from above and below for the IDOS of quantum systems on the entire spectrum. This landscape law, derived from the localization landscape, is not only much faster to compute than the entire IDOS, especially in random or disordered systems, but it also captures the scaling behavior of Anderson models near the bottom of the spectrum in full generality, for instance accounting for the logarithmic correction distinguishing the binary and uniform Anderson models. In one dimension, the bounds are so close that a single formula approximates the IDOS throughout the entire spectrum, with only a prefactor $C_5$ and a multiplicative shift on the energy $C_6$ (consistent with the $1+d/4$ formula found in Ref.~\cite{Arnold2019b}). In two dimensions, the bounds still provide a satisfactory approximation to the IDOS, but they cannot be merged into a single formula. Instead, one needs to adjust the prefactor from the bottom to the top of the spectrum. In summary, the landscape law promises to be a remarkable tool for investigating the properties of IDOS in many random or disordered potentials, with or without spatial correlations, not only theoretically but also numerically. In particular, it opens the perspective of accurately assessing the density of states in systems of very large sizes without having to compute any eigenvalues.

\begin{acknowledgments}
This work was supported by grants to several of the authors from the Simons Foundation (Grants No.~601937, D.N.A.; No.~601941, G.D.; No.~601944, M.F.; No.~563916, S.M) and, in part, by the DMS NSF No.~1839077. The authors would like to thank Christophe Texier for insightful discussions.
\end{acknowledgments}

\appendix*

\section{MATHEMATICAL PROOFS}

This Appendix intends to highlight some ideas behind the mathematical proof of~Eq.~\eqref{eq:inequalities} for tight-binding Hamiltonians $\hat H$ defined on a $d$-dimensional lattice.

Let us start with an argument which is quite far from the proof, but which nonetheless indicates the possibility of a connection between the minima of $1/u$ and the eigenvalues ($u$ being the localization landscape defined as the solution to $\hat{H}u=1$). 

To this end, we recall Lifschitz's original intuition behind the exponential nature of Lifschitz tails, the asymptotics of the IDOS near the edges of the spectrum (in our case, near zero).  Since the potential $V$ is non-negative, for $\hat H=-\Delta+V$ to have an eigenvalue smaller than some $E>0$, both $\ev{-\Delta}{\psi}$ (the kinetic energy) and $\ev{V}{\psi}$ (the potential energy) must be smaller than $E$. The eigenvalues of the Laplacian are, of course,  known explicitly, and one can see that for the first condition to be satisfied, $\psi$ has to be spread out on a cube of side length $r$ with $E\sim r^{-2}$. For $\ev{V}{\psi}$ to be below $E$ on such a cube, $V$ itself has to be smaller than $E$ on most of the sites. And, since the total number of sites in such a cube is $r^d \sim E^{-d/2}$, the probability that $V$ is desirably small is approximately $e^{-cr^d}\sim e^{-cE^{-d/2}}$. This suggests $N(E)\sim e^{-cE^{-d/2}}$ for the Anderson model, but for us it will carry a different information: a possibility of comparison to the landscape. 

Indeed, by the same token, what would it take to have $\min \frac 1u<E$? If we assume for the moment that $V=0$ on some cube $Q$ of side length $r$ and that $u$ has zero Dirichlet boundary data on its boundary, we observe that $\min_Q \frac 1u \sim r^{-2}$ (e.g., in 1D, we can write down the solution $u$ explicitly: $u=\frac 12 (i-i_0)(i_1-i)$ on $[i_0, i_1]$). A small $V$ would not distort this picture too much and hence, yet again, we discover that in order to have $\min_Q \frac 1u \sim E$ we need a cube of size $r^d \sim E^{-d/2}$ where the potential is small---exactly the same condition as the one which governs the eigenvalues according to Lifschitz's ideas. 

In terms of rigorous mathematics, there are many problems with this argument. First, neither the eigenvalues nor $u$ are local quantities, so a restriction to the Dirichlet problem on $Q$ is not fully justified. In the case of the eigenvalues, this problem is solved by the so-called Dirichlet-Neumann bracketing showing that the boundary values do not affect the situation too much. In the case of the landscape, the situation is more complicated: By the maximum principle the bigger the boundary data, the bigger the solution inside. Furthermore, such an argument could only provide a one-sided estimate: One needs different (and much more involved) considerations to assure that small eigenvalues cannot arise by any other mechanism than the one described above, and similarly for the landscape. Finally, and most importantly, all this mainly applies to small eigenvalues and asymptotic results. Nevertheless, it is an important insight which lets us set up the scales and justifies the definition of the counting function $N_u$ as the number of boxes of size $r^d \sim E^{-d/2}$ where $\min_Q \frac 1u \lesssim r^{-2}$.

Another important insight comes from a new version of the uncertainty principle guided by the landscape.  It was proved in Ref.~\cite{Arnold2019} that $\hat H$ is conjugate to a Hamiltonian which replaces the potential $V$ with $\frac 1u$:
\begin{align}\label{eq:uncert}
	\frac 1u \hat{H} u= -\frac{1}{u^2} \nabla\cdot  (u^2 \nabla) +\frac 1u, 
\end{align}
so 
\begin{align}\label{eq:E}
E &= \braket{\nabla \psi}{\nabla \psi} + \ev{V}{\psi} \nonumber\\
&= \braket{u\nabla \frac{\psi}{u}}{u\nabla \frac{\psi}{u}} + \ev{\frac 1u}{\psi}. 
\end{align}
The last inequality can be viewed as an uncertainty principle. Since the first term on the right-hand side is positive, it immediately implies that $E>\min \frac 1u$, providing yet another connection between the energies and the minima of $\frac 1u$, although, in fact, we will need the full identity to study the trade-off between the kinetic and potential energy governed by $u$, a clever localization procedure, and some fine properties of the landscape as a solution of an elliptic PDE to prove the inequalities (5) in the main paper. 

Let us now discuss the argument in a little more details. The most general form of the landscape law \cite{David2019landscape} establishes that for {\it all} potentials bounded from below
\begin{align}\label{eq:Law1}
	C_0\alpha^d N_u(C_1\alpha^{d+2}E)-&C_2N_u(C_3\alpha^{d+4} E) \nonumber \\
	&\le   N(E)\le N_u(C_4E), 
\end{align}
with the constants $C_i$ and some sufficiently small parameter $\alpha>0$ depending on the dimension only. There are no restrictions on the (probabilistic or deterministic) nature of the potential, or the size of the underlying domain, and there is no dependence of the constants on the energy $E$ or the potential. 

The polynomial correction of a lower bound is somewhat unpleasant, but irrelevant in most applications via a certain dichotomy. Indeed, if the behavior of $N(E)$ and $N_u(E)$ is exponential at the bottom of the spectrum (one of the signatures of localization, and a consequence of the aforementioned estimates for Lifschitz tails) then, of course, the polynomial corrector is irrelevant. Actually, for periodic potentials or for reasonably decaying ones (the so-called Kato class), the polynomial corrector disappears for different reasons. In either case, we ultimately get 
\begin{equation}\label{eq:inequalities-bis}
 C_5 N_u \left( C_6 E \right) \leq N(E) \leq N_u \left( C_4 E \right).
\end{equation}

Let us now say a few words about the proof of \eqref{eq:Law1}. If one can bound the energy for a collection of functions $\psi$ which do not necessarily represent eigenmodes but span some linear subspace, then the IDOS $N(E)$ can be estimated using the dimension (codimension) of this subspace. Clearly, we need this dimension to be related to $N_u$ in a desirable fashion. So the key is to pick an appropriate collection of cubes $Q$ and then an appropriate collection of functions $\psi$ to test our Hamiltonian. The upper bound on $N(E)$ is achieved by taking the collection of cubes $Q$ of side length approximately $E^{-1/2}$ such that $\min_Q \frac 1u \lesssim E$, and then considering functions $\psi$ whose average is zero on all cubes $Q$ as above. It can be shown that for all such functions $\ev{\hat H}{\psi} \geq E \braket{\psi}{\psi}$, and the choice of the collection of cubes assures the connection with $N_u$. 

To attack the lower bound \eqref{eq:Law1}, we start with a similar collection of cubes but now consider test functions $\psi$ given by a (smooth) restriction of the landscape $u$ to the cubes $Q$. One wants to show the opposite inequality, $\ev{\hat H}{\psi} \le E \braket{\psi}{\psi}$ in the corresponding subspace, and this is where the variation of $u$ on $Q$ becomes important: One needs some control on how nice the test function is. If $u^2$ is a doubling weight so its average on $Q$ is roughly speaking proportional to its average on $2Q$ (which is the case, for instance, when $V$ is periodic or rapidly decaying) then we can use some self-improvement properties for solutions of elliptic equations and  obtain the lower bound from \eqref{eq:inequalities-bis}. Here and everywhere by multiples of $Q$ we mean the cubes concentric with $Q$ with the side length multiples by the corresponding factor. If no such control on $u$ is \emph{a priori} known (it could be violated, for instance, for Anderson potentials), one has to be more careful and work with a collection of cubes $Q$ such that $\min_{Q_{\text{child}}}\frac 1u \leq \alpha^{d+2}\mu$ and  $\min_{Q}\frac 1u \geq \alpha^{d+4}\mu$, where $Q_{\text{child}}$ is concentric with $Q$ but has a much smaller side length. This ultimately yields a more complex lower bound \eqref{eq:Law1}.

Now to pass from \eqref{eq:Law1} to \eqref{eq:inequalities-bis} for Anderson potentials, one needs to show that $N_u(E)$ behaves exponentially at the bottom of the spectrum, so the polynomial correction could be absorbed. The arguments that we outlined in the beginning of this sketch are helpful, albeit  one-sided and too crude (we note that the actual precision of the exponential-type estimates needed for such an absorption supersedes even what has been known for classical Lifschitz tails). However, here we restrict our attention only to principal challenges, and perhaps it is fair to say that the biggest problem is the non-local character of $u$: even if $V$ is literally zero on a cube, the boundary data on $\partial Q$ could make $u$ itself arbitrarily small or large. 

It turns out that the solution  to the problem lies in the fact that the problem is bigger than anticipated. We show that not only the landscape can be large in $Q$ when  its boundary data on $\partial Q$ is large but, actually, if $u$ is large somewhere in $Q$ and $V$ is non-negligible, than there is a point on $\partial Q$ (or close to it) where $u$ is even larger. This result comes with quantitative estimates and, ultimately, allows one to set up a multiscale argument and conclude that $u$ blows up at the boundary of the entire domain, arriving at a contradiction. This shows that too large values of $u$ (equivalently, too small values of $1/u$) have the same probability as negligible values of $V$, that is, an exponentially small probability.  

To be a little more precise, we want to study the local behavior of $u$ on those cubes $Q$ of side length $r\sim E^{-1/2}$ where the potential has at least $\varepsilon$ portion of non-negligible values:
\begin{align}\label{eq:event}
	{\rm Card}\{i\in Q: V_i\gtrsim E\}\ge \varepsilon r^d.
\end{align}
This is exactly the setting in which $u$ being large deeply in the interior of $Q$ implies that $u$ on the boundary is even larger. Suppose $\min_Q\frac 1u \lesssim 1/r^2$, i.e., $ \max_Qu\gtrsim  r^2$. Let $(\sqrt{1+\varepsilon} \,r) Q\supset Q$ be a slightly larger cube of side length $\sqrt{1+\varepsilon} \,r$. Using some PDE considerations, starting from the subaveraging property for the Laplacian, one can show that whenever \eqref{eq:event} is true the maximum of $u$ on  $(\sqrt{1+\varepsilon} \,r) Q$ increases at least by a factor $(1+\varepsilon)$ compared to the values of $u$ in $Q/2$. One can gain some intuition looking at a precursor of this effect in 1D. Indeed $-u''+Vu=1$, which means that every time $u$ is large and $V$ is non-negligible, the second derivative of $u$ grows by this large amount and, hence, $u$ becomes substantially larger at least in one direction. If $V$ is negligible, for instance zero, then the second derivative drops  by 1, but overall the growth is overwhelming. Somewhat similar, although much more technical considerations apply in higher dimensions as well. In particular, one can find an $i'$ in the $\sqrt{1+\varepsilon}$-enlarged cube  such that $u_{i'}\ge  (1+ \varepsilon)r^2$. Inductively, if the next $\sqrt{1+\varepsilon}$-enlarged cube again has at least $\varepsilon$ portion of $V\gtrsim E$, one can repeat the above construction, and ultimately, we obtain a sequence of $i_k$ such that $u_{i_k}\gtrsim (1+\varepsilon)^kr^2\to \infty$ as $k\to \infty$. This leads to contradiction at infinity, or, if the domain is finite, to violation of the boundary data on the boundary of the entire domain. To prevent the contradiction, one has to exclude the events of the form \eqref{eq:event}, with growing side length $r_k\sim (1+\varepsilon)^kr$. The total probability of these events can be estimated by  $\sum_ke^{-cr^d_k}\sim e^{-cE^{-d/2}}$, which leads to the bound $N_u(E)\lesssim e^{-cE^{-d/2}}$.

In the other direction, \eqref{eq:uncert} implies that for cubes of side length $r\sim E^{-1/2}$,
\begin{equation}
\min_{Q}1/u\lesssim {\rm{ave}}_QV+E.
\end{equation}
Therefore, ${\mathbb P}(\min_{Q} 1/u\le  E)\ge {\mathbb P}({\rm{ave}}_QV \lesssim E)$, which can also be bounded from below by $e^{-cE^{-d/2}}$. All in all, we get exponential bounds for $N_u(E)$ near $E=0$ which allows us to self-improve \eqref{eq:inequalities-bis} to \eqref{eq:Law1}, as desired.

\bibliographystyle{apsrev4-1}
\bibliography{bibliography.bib}

\end{document}